\documentclass[a4paper,11pt]{article}
\pdfoutput=1 

\usepackage{jheppub} 
\usepackage{bbm,bm,graphicx,mathtools,color,slashed,hyperref}
\usepackage{amssymb,amsmath}
\usepackage{dsfont}
\usepackage{physics}

\newcommand{\diff}{\mathrm{d}}

\newcommand{\im}{\mathrm{i}}

\newcommand{\rme}{\mathrm{e}}

\newcommand{\khat}{\hat{k}}

\title{Lattice gradient flows (de-)stabilizing topological sectors}

\author[1]{Yuya Tanizaki,}
\affiliation[1]{Yukawa Institute for Theoretical Physics,
Kyoto University, Kyoto, 606-8502, Japan}
\emailAdd{yuya.tanizaki@yukawa.kyoto-u.ac.jp}

\author[2,3]{ Akio Tomiya,}
\affiliation[2]{Department of Information and Mathematical Sciences,
Tokyo Woman’s Christian  University, Tokyo 167-8585, Japan}
\affiliation[3]{RIKEN Center for Computational Science,
Kobe 650-0047, Japan}
\emailAdd{akio@yukawa.kyoto-u.ac.jp}

\author[1]{ Hiromasa Watanabe}
\emailAdd{hiromasa.watanabe@yukawa.kyoto-u.ac.jp}

\preprint{YITP-24-159}

\abstract{
We investigate the stability of topological charge under gradient flow taking the admissibility condition into account. 
For the $SU(2)$ Wilson gauge theory with $\beta=2.45$ and $L^4=12^4$, we numerically show that the gradient flows with the Iwasaki and DBW2 gauge actions stabilize the topological sectors significantly, and they have qualitatively different behaviors compared with the Wilson and tree-level Symanzik flows. 
By considering the classical continuum limit of the flow actions, we discuss that the coefficient of dimension-$6$ operators has to be positive for stabilizing the one-instanton configuration, and the Iwasaki and DBW2 actions satisfy this criterion while the Wilson and Symanzik actions do not. 
Moreover, we observe that the DBW2 flow stabilizes the topological sectors at the very early stage of the flow ($\hat{t}\approx 0.5$--$1$), suggesting that a further systematic investigation of the DBW2 flow is warranted to confirm its computational efficiency in determining the gauge topology.
}

\begin{document}

\maketitle

\section{Introduction}

Gauge fields with nontrivial topology play crucial roles in understanding nonperturbative phenomena such as confinement, chiral symmetry breaking, etc., which are fundamental features of quantum chromodynamics (QCD). 
Moreover, the property of the anomalously-broken $U(1)_A$ symmetry significantly affects the finite-temperature QCD phase diagram in the chiral limit~\cite{Pisarski:1983ms}, and it is crucial to understand gauge topologies correctly for uncovering those phenomena in the first-principle nonperturbative computations.

Despite this fact, the standard lattice regularization of gauge theories spoils the topology completely because the space of gauge fields becomes $SU(N)^{\otimes (\mathrm{links})}$, which is a connected space. 
This apparent mismatch can be resolved by noticing that the continuum limit corresponds to the weak-coupling limit for asymptotically free theories~\cite{Luscher:1981zq}. In the weak-coupling limit, the path integral should localize on the field configurations with $||\bm{1}-U_{x,\mu\nu}||^2\lesssim  O(1/\beta)$. As a result, one can set the admissibility condition for the lattice fields, $\sup_{x,\mu\nu}||\bm{1}-U_{x,\mu\nu}||^2\ll 1$, without affecting the continuum limit. The topological sector becomes well-defined once we impose the admissibility constraint~\cite{Luscher:1981zq, Phillips:1986qd}. 

This resolves the problem of the gauge topology at the foundational level of lattice gauge theories. However, imposing the admissibility condition requires sufficiently weak lattice gauge couplings ($\beta \gtrsim (
\text{Admissibility bound})^{-1} \approx 30$), and we have to prepare a huge lattice size, which is impractical for numerical investigations with the current computational power. 
As a practical tool to define (or ``assign'') a gauge topology, people use some cooling algorithms to obtain smooth gauge fields out of rough lattice gauge fields~\cite{Teper:1985rb, Hoek:1986nd}. 
After such manipulations, its topology can be directly calculated by taking the lattice sum of the topological charge density, $\frac{\varepsilon_{\mu\nu\rho\sigma}}{32\pi^2}\tr(F_{\mu\nu}F_{\rho\sigma})$, or counting the zero modes of the lattice Dirac operator.
There are various proposals for cooling algorithms, but they can be systematically understood as the approximation of the gradient flow~\cite{Luscher:2010iy}. Therefore, let us focus on the gradient flow as the cooling procedure in this paper. 

The gradient flow for lattice gauge fields takes the form of the heat equation in the continuum limit, and thus it smears the fields over the size $\sqrt{8t}$, where $t$ is the flow time. 
To define the topology, we want to eliminate the lattice-scale fluctuations, so it is natural to require $\sqrt{8t}\gg 2a$. 
However, when should we stop the flow? 
The common lore states that one should choose a ``reasonable'' $t$ that satisfies 
\begin{equation}
    2a\ll \sqrt{8t}\ll \frac{La}{2}, 
    \label{eq:conventional_bound}
\end{equation}
where $L$ is the number of lattice sites along the shortest direction. When $\sqrt{8t}\approx La/2$, the flow starting from a single-lattice point meets at the opposite of the lattice by wrapping around the spacetime geometry due to the periodic boundary condition, and the locality is severely lost. 
Therefore, the bound \eqref{eq:conventional_bound} is reasonable to measure the correlation functions of generic local operators. 
To find a reasonable value of $t$, we have to look at the data carefully and choose it \textit{a posteriori} in the actual numerical simulation. 

In this paper, we discuss the (in-)stability of the topological sectors under gradient flows, so we shall focus on the integrated topological charge and its susceptibility. 
In the continuum limit, the topological charge never changes under the continuous deformation of gauge fields: However long we apply the gradient flow, we should get the same result. 
If this continuum intuition were correct, we do not need to care about the conventional upper bound \eqref{eq:conventional_bound} for those topological quantities. We shall investigate if we can realize this intuition with the finite lattice spacing. 

To this end, we should ensure that the gradient flow maintains the topological sector for $\sqrt{8t}\gg 2a$. 
As a necessary condition for this property, a one-instanton configuration has to be stable under the gradient flow. However, it has been known for a long time~\cite{Iwasaki:1983bv, Iwasaki:1983iya} that the stability of instanton depends on the detail for the lattice action of the gradient flow. For example, the most common choice, the Wilson flow, eliminates all the instantons in the long flow-time limit. 
As we will see analytically in Section~\ref{sec:AnalyticalStudy}, we can stabilize the one-instanton configuration under the lattice gradient flow by taking the ``over-improved'' flow action~\cite{GarciaPerez:1993lic, deForcrand:1995bq}. 
We shall test if this stability of the one-instanton configuration also practically implies the stability of topological sectors for the gauge configurations of actual Monte Carlo simulations.

This study compares four kinds of gradient flows, Wilson, Symanzik, Iwasaki, and DBW2, as the lattice flow action, which only consists of the plaquette and rectangular terms (Section~\ref{sec:ChoiceofFlow}). According to the analytical study, Iwasaki and DBW2 flows stabilize one-instanton configurations, while Wilson and Symanzik flows eliminate it, and we numerically confirm it in Section~\ref{sec:Numerics_OneInstanton}. 

In Section~\ref{sec:Numerics_MonteCarlo}, we apply these gradient flows for the Monte Carlo configurations of the $SU(2)$ Wilson gauge theory with $\beta=2.45$ and the lattice size $L^4=12^4$. 
Along the gradient flow, we not only measure the topological charge and the Wilson action but also monitor the maximal value of the local field strength, which is the measure for the admissibility condition. 
We apply the gradient flows until $t/a^2=50$, which largely exceeds the conventional bound $t/a^2\lesssim L^2/32=4.5$. We find that both Iwasaki and DBW2 flows achieve the stability of topological sectors also for Monte Carlo configurations, and the topological susceptibility stays almost constant. As a byproduct of this long-time stability, we find lattice configurations that numerically satisfy the Bogomol'nyi–Prasad–Sommerfield (BPS) bound for various topological sectors ($|Q_{\mathrm{top}}|\le 5$). 
Wilson and Symanzik flows do not have this property, and the topological susceptibility gradually decreases along the flow. 
It also turns out that Iwasaki and DBW2 flows have a better property on the short-time behavior: The topological charges converge to integers quickly, while topological charges obtained by Wilson and Symanzik flows scattered to non-integer values at short flow times.
We mention the summary and give discussions in Section~\ref{sec:Summary}.

\section{Lattice gradient flows and stability of one-instanton configurations}
\label{sec:AnalyticalStudy}

We consider the hypercubic lattice and assume that we want to assign the topological charge for the link variables $\{U_{x,\mu}\}$ using the gradient flow: 
\begin{equation}
    \frac{\diff }{\diff t}U_{x,\mu}(t)=-\frac{\delta}{\delta U_{x,\mu} }S_{\mathrm{flow}}[U_{x,\mu}(t)], 
\end{equation}
where $U_{x,\mu}(t=0)=U_{x,\mu}$. 
$S_{\mathrm{flow}}$ defines the height function on the space of link variables, and it monotonically decreases along the gradient flow until it hits a classical solution.

\subsection{Properties of the flow action in the classical continuum limit}
\label{sec:AnalyticalStudy_ClassicalContinuum}

In this paper, we focus on the flow action $S_{\mathrm{flow}}$ that consists of the plaquette and rectangular terms, 
\begin{align}
    S_{\mathrm{flow}}&=c_0 S^{(\mathrm{plaq.})} + 2 c_1 S^{(\mathrm{rect.})}, 
    \label{eq:FlowAction}
\end{align}
where 
\begin{align}
    S^{(\mathrm{plaq.})}&=\sum_{x,\mu\neq\nu}\mathrm{Re}\bigl[\tr(\bm{1}-\underbrace{U_{x,\mu}U_{x+\hat{\mu},\nu}U^\dagger_{x+\hat{\nu},\mu}U^\dagger_{x,\nu}}_{=:\, U_{x,\mu\nu}^{(1\times 1)}=\, U_{x,\mu\nu}})\bigr], 
    \label{eq:Plaquette}\\
    S^{(\mathrm{rect.})}&=\sum_{x,\mu\neq\nu}\mathrm{Re}\bigl[\tr(\bm{1}-\underbrace{U_{x,\mu}U_{x+\hat{\mu},\mu}U_{x+2\hat{\mu},\nu}U^\dagger_{x+\hat{\mu}+\hat{\nu},\mu}U^\dagger_{x+\hat{\nu},\mu}U^\dagger_{x,\nu}}_{=:\, U^{(2\times 1)}_{x,\mu\nu}})\bigr]. 
    \label{eq:Rectangular}
\end{align}

Let us take the classical continuum limit to understand the property of the flow action. We then assume that the link variable is given by 
\begin{align}
    U_{x,\mu}=\mathcal{P}\exp\left[\im a \int_0^1 \diff s \, A_\mu(x + s a \hat{\mu})\right], 
\end{align}
where $A_\mu(x)$ is the $SU(N)$ gauge field on the continuum spacetime. 
We here introduce the lattice spacing $a$ as the formal power-counting parameter, and the lattice sites are located as $x\in (a\mathbb{Z})^4$. 
Then, the flow action can be expanded in terms of $a$ as~\cite{Luscher:1984xn, GarciaPerez:1993lic} 
\begin{align}
    S_{\mathrm{flow}}=(c_0+8c_1)\frac{a^4}{2}\sum_{x,\mu\nu}\tr(F_{\mu\nu}^2)-(c_0+20 c_1)\frac{a^6}{12}\sum_{x,\mu\nu}\tr[(D_\mu F_{\mu\nu})^2] + O(a^8). 
    \label{eq:FlowAction_continuum}
\end{align}
The first term corresponds to the continuum Yang--Mills action after replacing $a^4\sum_x$ by $\int \diff^4 x$, and the second term gives the leading lattice artifact. 
To find this expression, it is convenient to note that $\sum_{\mu\nu}\tr(F_{\mu\nu}^2)$ and $\sum_{\mu\nu}\tr[(D_\mu F_{\mu\nu})^2]$ are the unique dimension-$4$ and -$6$ operators, respectively, that can come out of the expansion of the plaquette and rectangular terms.\footnote{For the dimension-$4$ operators, $\sum_{\mu\nu}\tr(F_{\mu\nu}^2)$ is the only gauge-invariant term invariant under both hypercubic rotation and parity. The dimension-$5$ operators cannot appear as it violates parity. The hypercubic and parity invariance does not select $\sum_{\mu\nu}\tr[(D_\mu F_{\mu\nu})^2]$ for dimension-$6$ operators, but an expansion of the plaquette and rectangular terms can only produce the operators living in the $2$-dimensional $\mu\nu$ plane. As a result, $\sum_{\mu\nu}\tr[(D_\mu F_{\mu\nu})^2]$ is the unique possibility.} 
Then, we may take the Abelian gauge field to replace the path-ordered exponential with the usual exponential as the $SU(N)$ covariantization does not have any ambiguity.\footnote{For dimension-$8$ operators, this Abelian trick does not work since the covariantization of $\partial_\mu \partial_\nu F_{\mu\nu}$ can be either $D_\mu D_\nu F_{\mu\nu}$ or $D_\nu D_\mu F_{\mu\nu}$, which has a difference by $F_{\mu\nu}^2$. } 
For Abelian gauge fields, the $n_1\times n_2$ loop $U^{(n_1\times n_2)}_{x,\mu\nu}$ can be evaluated as $\exp[\im a^2\int_0^{n_1}\diff s_1 \int_0^{n_2}\diff s_2\, F_{\mu\nu}(x+a s_1 \hat{\mu} + a s_2 \hat{\nu})]$, and we obtain \eqref{eq:FlowAction_continuum}. 

In the following, we set 
\begin{align}
    c_0=1-8c_1, 
\end{align}
so that the coefficient of the classical Yang--Mills term becomes $1$. 

\subsection{Stability of the one-instanton configuration}
\label{sec:AnalyticalStudy_StabilityInstanton}

In this section, we discuss the stability of the one-instanton gauge configuration under the gradient flow~\cite{GarciaPerez:1993lic}. 
We here assume that the four-dimensional lattice is infinitely large and the size of instanton $\rho$ is much larger than the lattice scale $a$: $\rho \gg a$. 

Let us evaluate the flow action for the $SU(2)$ one-instanton configuration~\cite{Belavin:1975fg, tHooft:1976snw}, 
\begin{align}
    A^{\mathrm{inst}}_\mu(x;x_*,\rho)= -\sum_{\nu}\frac{\eta_{\mu\nu}(x_\nu-x_{*,\nu})}{(x-x_*)^2+\rho^2},
    \label{eq:InstantonConfiguration}
\end{align}
where $\eta_{\mu\nu}=\sum_i \eta_{i\mu\nu}\tau_i$ is the self-dual 't~Hooft symbol given by $\eta_{i\mu\nu}=\varepsilon_{i\mu\nu4}+\delta_{i\mu}\delta_{\nu4}-\delta_{i\nu}\delta_{\mu 4}$ and Pauli matrices $\tau_i$, and $x_*$ is the location of the instanton. The field strength is then given by 
\begin{equation}
    F^{\mathrm{inst}}_{\mu\nu}(x; x_*,\rho)=\frac{2\rho^2}{[(x-x_*)^2+\rho^2]^2}\eta_{\mu\nu}, 
\end{equation}
which satisfies the self-dual Yang--Mills equation thanks to $\eta_{\mu\nu}=\frac{1}{2}\sum_{\rho\sigma}\varepsilon_{\mu\nu\rho\sigma}\eta_{\rho\sigma}$. 

Let us now substitute the instanton \eqref{eq:InstantonConfiguration} into the flow action~\eqref{eq:FlowAction_continuum}. Evaluating the lattice sum $a^4 \sum_x$ is cumbersome, but we may replace it with the integral $\int \diff^4 x$ up to an exponentially small correction for $\rho\gg a$~\cite{GarciaPerez:1993lic}. 
Just for simplicity, let us demonstrate this fact in the one-dimensional lattice sum: Assume that $f(x)$ is a smooth function over the real axis, then the one-dimensional lattice sum can be expressed using the complex contour integral as
\begin{align}
    a\sum_{x\in a\mathbb{Z}} f(x)&=\oint \diff x \frac{1}{\rme^{\im \frac{2\pi}{a}x}-1}f(x) \notag\\
    &=\int_{-\infty}^{\infty}\diff x\left[\frac{1}{\rme^{\frac{2\pi \im}{a}(x-\im 0^+)}-1}f(x)-\frac{1}{\rme^{\frac{2\pi\im}{a}(-x+\im 0^+)}-1}f(-x)\right] \notag\\
    &=\int f(x)\diff x + \int_{-\infty}^{\infty}\diff x\, \frac{f(x)+f(-x)}{\rme^{\frac{2\pi \im}{a}(x-\im 0^+)}-1}.  
\end{align}
The second term of the last line gives the finite-$a$ correction for the replacement. We can evaluate it by closing the contour on the lower half plane, and for the instanton of the size $\rho$, the singularity of the integrand on the lower-half plane should be around $x=\pm x_*-\im \rho$. Thus, the finite-$a$ correction due to the replacement $a^4 \sum_x \Rightarrow \int \diff^4x$ is controlled as $O(\rme^{-2\pi \rho/a})$ when $\rho\gg a$. 
As a result, the flow action for $A^{\mathrm{inst}}_\mu$ is given by\footnote{For the $O(a^2)$ term, we use $D_\mu F^{\mathrm{inst}}_{\mu\nu}=-\frac{4\rho^2}{(x^2+\rho^2)^3}[3\eta_{\mu\nu}x_\mu-(1-\delta_{\mu\nu})\sum_\rho \eta_{\rho \nu}x_\rho]$ (No sum over $\mu$). }
\begin{align}
    S_{\mathrm{flow}}[A^{\mathrm{inst}}_{\mu}(x;x_*,\rho)] 
    &= \int \diff^4 x \sum_{\mu\neq\nu} \left\{\frac{1}{2}\tr[(F^{\mathrm{inst}}_{\mu\nu})^2]-\frac{(1+12c_1)a^2}{12}\tr[(D_\mu F^{\mathrm{inst}}_{\mu\nu})^2] +O(a^4)\right\}\notag\\
    &=8\pi^2\left[1- \frac{1+12 c_1}{5} \left(\frac{a}{\rho}\right)^2 + \cdots\right]. 
\end{align}
This clarifies that the property of the lattice gradient flow has the important qualitative difference between $1+12 c_1 >0$ and $1+12 c_1<0$. 
To our best knowledge, this fact was first numerically investigated essentially in \cite{Iwasaki:1983bv, Iwasaki:1983iya} and later more systematically understood in~\cite{GarciaPerez:1993lic}.\footnote{In~\cite{GarciaPerez:1993lic, deForcrand:1995bq}, the idea of flipping the sign for the dimension-$6$ operator is called ``over-improvement.'' They realize the over-improvement by combining the $1\times 1$ and $2\times 2$ plaquettes in the numerical computation, while \cite{GarciaPerez:1993lic} also discusses the rectangular term for analytical purpose. } 

\subsubsection*{Instability of one-instanton for $c_0+20c_1=1+12 c_1 \ge 0$}

Let us consider the case $1+12 c_1>0$, and then the coefficient of the dimension-$6$ operator in \eqref{eq:FlowAction_continuum} is negative. 
In the continuum, the size of the instanton is a moduli parameter, but the flow action gets smaller due to the finite-$a$ correction for smaller instantons when $\rho\gg a$. 
As a result, the size of instanton gradually shrinks along the gradient flow with $1+12 c_1>0$, and it eventually becomes comparable to the lattice scale $a$. 
Although the properties of the flow for $\rho\sim a$ are not easy to know in an analytical way, it is natural to imagine that such small instantons slip away through holes of the lattice.\footnote{It is sometimes claimed that the over-smearing of the topological charge is an infrared (IR) effect, but this should be understood with the great care. 
When taking it literally in a naive way, one may think that the gauge field spreads too much and becomes dilute as the effect of smearing, and the instanton disappears. 
What actually happens is the opposite as we have seen in the main text. We also note that the disappearance of instanton is not caused by the wrap-around effect as it also happens in the infinite lattice. Due to the presence of the size moduli for instantons, the classical action stays constant even if the field strength gets more and more localized. The nonlinear IR effect of the lattice gradient flow causes this localization for one-instanton configuration when $1+12 c_1>0$. When the gauge field becomes localized on a few lattice sites, we can no longer discriminate if we have an instanton or just a lattice artifact. } 

Let us state the same thing differently. When $\rho\gg a$, the plaquette at each site is close to the identity and satisfies the admissibility condition: Let us define the matrix norm $||A||$ as the maximal eigenvalue of $\sqrt{A^\dagger A}$, and 
\begin{align}
    \sup_{x,\mu\nu}||\bm{1}-U_{x,\mu\nu}||^2 \simeq \sup_{x,\mu\nu} a^4||F_{\mu\nu}(x)||^2\le \frac{4a^4}{\rho^4} \ll 1.
\end{align}
As L\"uscher pointed out~\cite{Luscher:1981zq}, the $SU(N)$ gauge bundle can be constructed for admissible gauge fields and the topological charge can be defined: Any continuous deformation of lattice gauge fields maintains the topological charge as long as the admissibility is kept intact. 
However, the above argument suggests that the gradient flow with $1+12 c_1>0$ eliminates the one-instanton configuration, and the topological charge jumps from $1$ to $0$. 
Even if we start the gradient flow from admissible gauge fields, the flowed configuration violates the admissibility. 

For $1+12 c_1=0$, the leading lattice artifact vanishes, and we have to look at the dimension-$8$ operator. 
This is computed in~\cite{GarciaPerez:1993lic}, and the $O((a/\rho)^4)$ term is negative for one-instanton, which again prefers smaller instantons. 
Therefore, for $1+12 c_1 \ge 0$, the one-instanton configuration is unstable under the gradient flow. 

\subsubsection*{Stability of one-instanton for $c_0+20c_1=1+12 c_1<0$}

For $1+12 c_1<0$, the property of the gradient flow is dramatically changed as the flow action for one-instanton prefers larger instantons. 
Therefore, if we start the lattice gradient flow with $\rho\gg a$, the instanton never disappears even with the infinite-time limit of the flow. 
If we perform the gradient flow for the finite lattice, it is easy to imagine that $\rho$ becomes comparable to the lattice size $L$, and the gradient flow effectively stops with the maximal field strength of the order of $1/L^2$, which obviously satisfies the admissibility criterion. 

The discussion here ensures that the gradient flow with $1+12 c_1<0$ respects the admissibility condition for the one-instanton configuration with $\rho \gg a$, and we will numerically confirm it in Section~\ref{sec:Numerics_OneInstanton}. 
A more interesting question is what happens if we apply the gradient flow for more generic configurations appearing in the Monte Carlo sampling. 
If the configuration consists of a dilute gas of large instantons, the gradient flow with $1+12 c_1<0$ will respect admissibility as this is the case for each isolated instanton. 
However, the confinement physics requires excitation of various magnetic objects, such as monopoles and center vortices~\cite{Nambu:1974zg, Mandelstam:1974pi, Polyakov:1975rs, tHooft:1977nqb, Cornwall:1979hz, Nielsen:1979xu, Ambjorn:1980ms}. For example, the intersection points of center-vortex worldsheets carry $1/N$ fractional topological charge~\cite{Engelhardt:1999xw, Reinhardt:2001kf, Cornwall:1999xw}, and they play a pivotal role in semiclassically understanding the large-$N$ behaviors of the topological susceptibility and its relation to the Witten--Veneziano-type $\eta'$ mass~\cite{Cornwall:1999xw, Tanizaki:2022ngt, Hayashi:2024qkm, Hayashi:2024yjc}.  It is highly nontrivial how the gradient flow acts on the ensemble of those objects, and it is important to investigate behaviors of different gradient flows in the actual gauge configurations of the Monte Carlo simulation, which will be addressed in Section~\ref{sec:Numerics_MonteCarlo}.

\section{Choice of the flow action}
\label{sec:ChoiceofFlow}

In our numerical study, we use the following four choices of $c_1$ for the gradient flow: Wilson ($c_1=0$), Symanzik ($c_1=-1/12$), Iwasaki ($c_1=-0.331$), and DBW2 ($c_1=-1.4088$). For Wilson and Symanzik flows, the one-instanton configuration shrinks, and it is unstable in the long flow-time limit even if its original size is large. 
For Iwasaki and DBW2 flows, the one-instanton configuration should be stable if its size is large enough. We list the information of these flows in Table~\ref{tab:coefratio}. 

\begin{table}[htb]
\centering
\begin{tabular}{c|l|l||l} 
Flow action & $c_0=1-8c_1$ & $c_1$ & $1+12 c_1 $ 
\\ \hline\hline
Wilson & $\phantom{1}1$ & $\phantom{-}0$ & $\phantom{-1}1.0$ \\
tree-level Symanzik & $\phantom{1}5/3$ &  $-1/12$ & $\phantom{-1}0.0$ \\
Iwasaki & $\phantom{1}3.648$ & $-0.331$ & \phantom{1}$-2.972$ \\
DBW2 & $12.2704$ & $-1.4088$ & $-15.9056$\\
\end{tabular}
\caption{
Coefficients in the flow actions.
The coefficient of the plaquette term is set as $c_0 = 1-8c_1$.
$1+12 c_1$ is a measure of stability for one-instanton configuration.
\label{tab:coefratio}
}
\end{table}

When used for the Boltzmann weight, these actions are well-established as improved actions with two-dimensional parameter space (see \cite{Necco:2003jf} for details).
Let us make a brief comment on this point taking a detour. 
In lattice simulations, 
the most standard choice is the Wilson plaquette action (i.e. $c_1=0$).
This has been used not only for the production of configurations but also for the gradient flow.
However, there is a freedom to change the lattice action by higher-dimensional operators, and one can use this ambiguity to achieve milder lattice artifacts.
One of the most common choices for improved gauge action is tree-level improved Symanzik gauge action (i.e. $c_1=-1/12$), which eliminates the dimension-$6$ operator.

The other two actions, Iwasaki and DBW2, correspond to renormalization-group (RG) improved actions. 
Any point in the renormalized trajectory of the RG flow has the same long-distance behavior, and such actions are called perfect actions~\cite{Hasenfratz:1993sp,DeGrand:1996zb,Hasenfratz:1998gu}. 
Iwasaki performed a block-spin RG transformation for $SU(N)$ gauge theory near the non-trivial fixed point and found an improved lattice gauge action~\cite{Iwasaki:1983iya}.
After that, QCD-TARO collaboration investigated the RG trajectory using the Schwinger--Dyson method \cite{QCD-TARO:1998nbk} and developed DBW2 (\underline{D}oubly \underline{B}locked from \underline{W}ilson action in \underline{two} coupling space \cite{QCD-TARO:1999mox}) as an approximation of the perfect action.\footnote{There are other improved actions beyond the two-coupling space $c_{0,1}$, such as the Zeuthen flow, which eliminates all $\mathrm{O}(a^2)$ artifacts for the flowed energy density \cite{Ramos:2015baa}. A similar study has been done with a fixed point action using machine learning  \cite{Holland:2024muu} in the same spirit of the original work \cite{Hasenfratz:1993sp}.
However, we omit them in this work because we here focus on the flows with $c_0=1-8c_1$ and $c_1$.}
Properties of RG-improved actions are studied in \cite{Necco:2003vh}. 
There is also a detailed study to compare these gauge actions on improving chiral properties of lattice fermions~\cite{DeGrand:2002vu}. 

A comparison of the definition of flows and cooling has been done in~\cite{Bonati:2014tqa,Alexandrou:2015yba,Alexandrou:2017hqw}. In this work, we compare topological charges of different flows with longer flow time taking into account the admissibility.
In Appendix~\ref{app:diffusion_length}, we discuss the diffusion length of these flows and confirm that it is roughly given by $\sqrt{8t}$ for any choice of $c_1$.

\section{Numerical analysis}
\label{sec:NumericalAnalysis}

In this section, we analyze the property of gradient flows listed in Table~\ref{tab:coefratio} numerically and perform two experiments with the $SU(2)$ gauge group. 
In Section~\ref{sec:Numerics_OneInstanton}, we apply gradient flows on a one-instanton solution on the lattice constructed by \cite{Fox:1985wp} and confirm the analytical prediction discussed in Section~\ref{sec:AnalyticalStudy_StabilityInstanton}.
Next, in Section~\ref{sec:Numerics_MonteCarlo}, we apply gradient flows for the gauge configurations generated by the Monte Carlo simulation of $SU(2)$ pure Yang--Mills theory.
All gradient flows are implemented by using the third-order Runge--Kutta method (RK3) with the RK time step $\epsilon = 0.01$
\cite{Luscher:2010iy}. 
Our numerical code is developed based on \texttt{Gaugefields.jl} in JuliaQCD \cite{Nagai:2024yaf}.

In our numerical study, we measure the lattice topological charge, which is expressed in the continuum notation as, 
\begin{align}
    Q_{\mathrm{top}}=a^4\sum_x \frac{1}{32\pi^2}\varepsilon_{\mu\nu\rho\sigma} \tr(F_{\mu\nu}F_{\rho\sigma}) + \text{lattice artifact},
\end{align}
with the $O(a^2)$ improvement by combining the clover-leaf term and clover-like rectangular terms to eliminate the dimension-$6$ contribution from the lattice artifact (see \cite{Alexandrou:2017hqw} for details).
In this work, we measure an indicator of the admissibility condition~\cite{Luscher:1981zq}. The matrix norm becomes simpler for the $SU(2)$ case, and let us use the following quantity:
\begin{align}
P_\mathrm{ adm} &= \frac{1}{2}\max_{x,\mu\nu} ||\bm{1}-U_{x,\mu\nu}||^2 \notag\\
&= \frac{1}{2}\max_{x, \mu\nu} \left|\tr (\bm{1}-U_{x,\mu\nu}) \right|.
\label{eq:adm}
\end{align}
The possible range of $P_{\mathrm{adm}}$ is $0\le P_{\mathrm{adm}}\le 2$. 
L\"uscher claimed $P_\mathrm{ adm} < 0.015$ is a non-optimal admissible bound~\cite{Luscher:1981zq},\footnote{We here say that the admissibility bound is $\varepsilon$ if the topological charge stays constant under any continuous deformations of the link variables satisfying $P_{\mathrm{adm}}<\varepsilon$. L\"uscher claimed that $\varepsilon=0.015$ can be chosen for $SU(2)$ to satisfy this property, but the possible maximal value for $\varepsilon$ is not known.  } and we show this bound in the figures for $P_{\mathrm{adm}}$ as a reference.

\subsection{Flows on one-instanton configurations}
\label{sec:Numerics_OneInstanton}

In this subsection, we discuss the (in-)stability of the one-instanton configuration under gradient flows. 
As the one-instanton configuration on the finite periodic lattice, we prepare the gauge link variables following~\cite{Fox:1985wp}, which contain one instanton with different radii $R=\rho/a$ for the lattice size $L=12$.\footnote{In \cite{Fox:1985wp}, it was numerically confirmed that their construction gives the configuration with the topological charge $1$ according to L\"uscher's geometric definition for $R=0.5$, $1.0$, $1.5$, $2.0$ with $L=6$.}

\begin{figure}[t]
\centering
\includegraphics[width=0.48\textwidth]{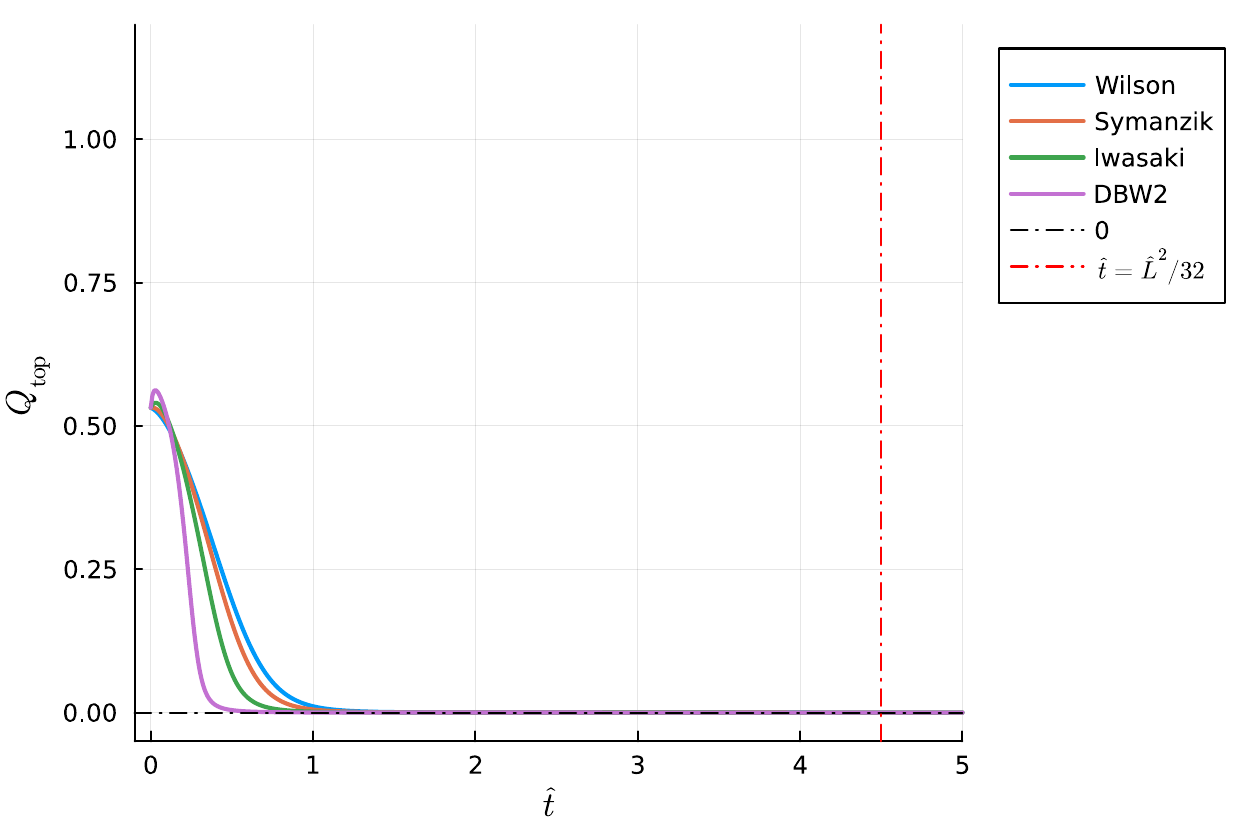}
\includegraphics[width=0.48\textwidth]{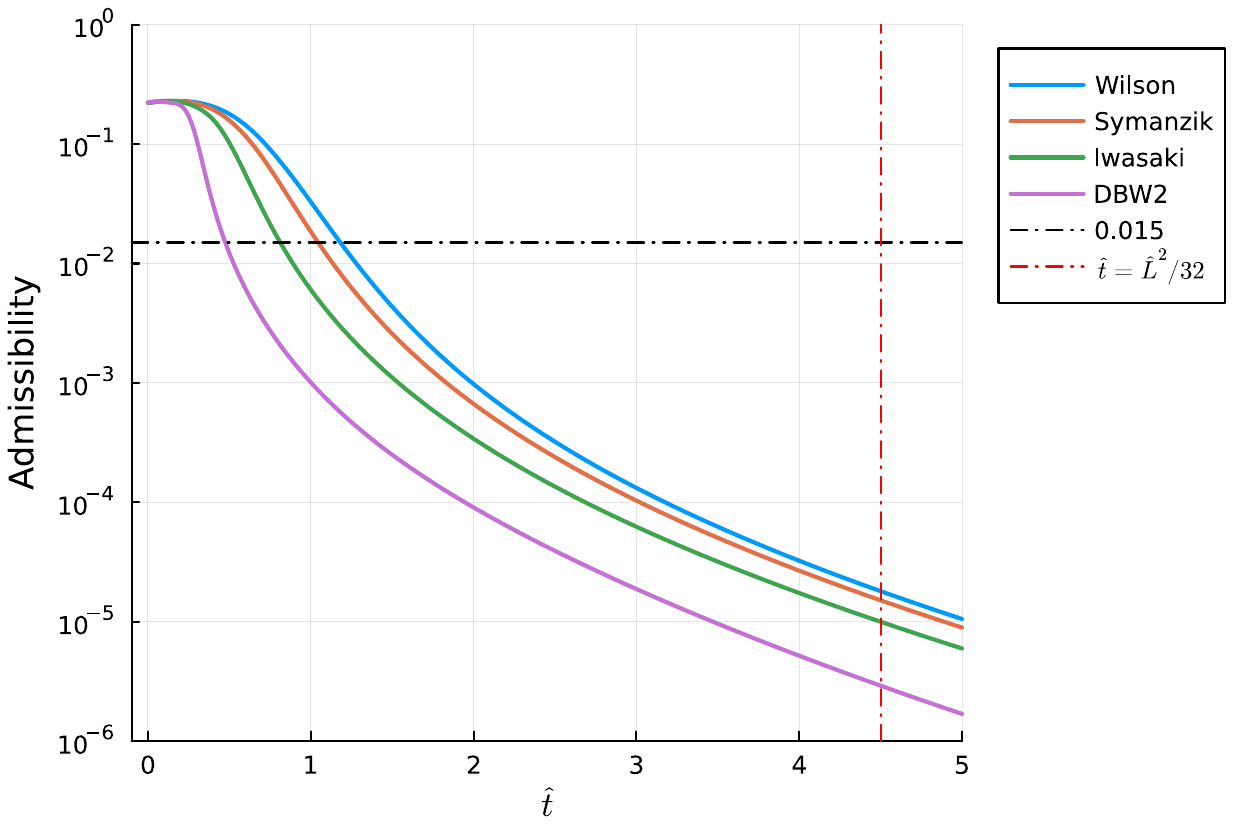}
\caption{
Topological charge and the admissibility under flows for an instanton with $R = 1.3$.
The horizontal axis is dimensionless flow time $\hat{t}=t/a^2$ and the vertical axis is the value of $Q_{\rm top}$ or $P_{\mathrm{adm}}$.
None of the flow can detect the instanton.
\label{fig:flow_with_instanton_R=1.3}
}
\end{figure}

Figure~\ref{fig:flow_with_instanton_R=1.3} shows the results for the topological charge $Q_{\mathrm{top}}$ and the admissibility measure $P_{\mathrm{adm}}$ for an instanton with radius $R=1.3$. 
The flow time is taken as $0\le \hat{t}=t/a^2 \le 5$. 
The instanton with $R=1.3$ is too small, and none of the flows can capture non-zero topological charges.
As we gradually make the instanton larger, those flows start to detect its nonzero topological charge. 
Among four gradient flows in Table~\ref{tab:coefratio}, DBW2 detects the smallest instantons with $R\approx 1.5$.

\begin{figure}[t]
\centering
\fbox{\begin{minipage}[b]{0.48\columnwidth}
    \includegraphics[width=1.0\columnwidth]{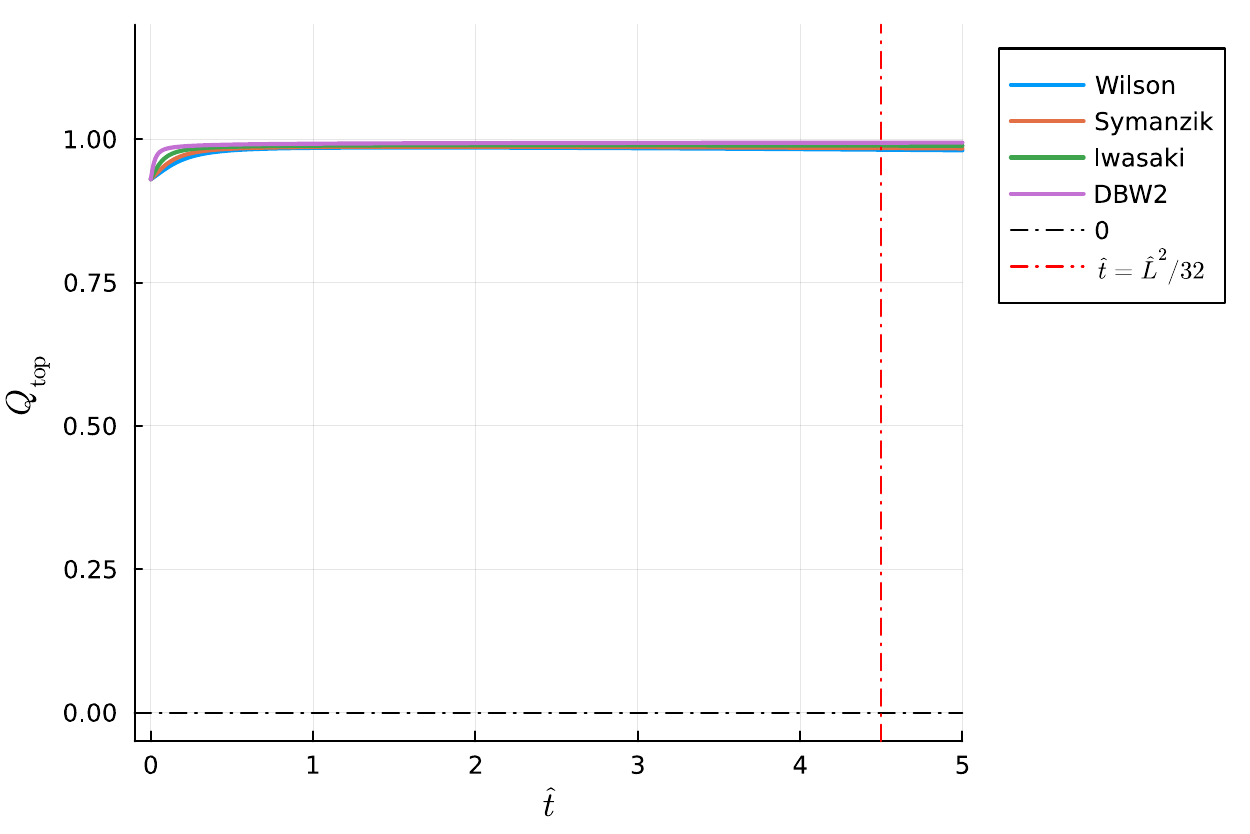}\\
    \includegraphics[width=1.0\columnwidth]{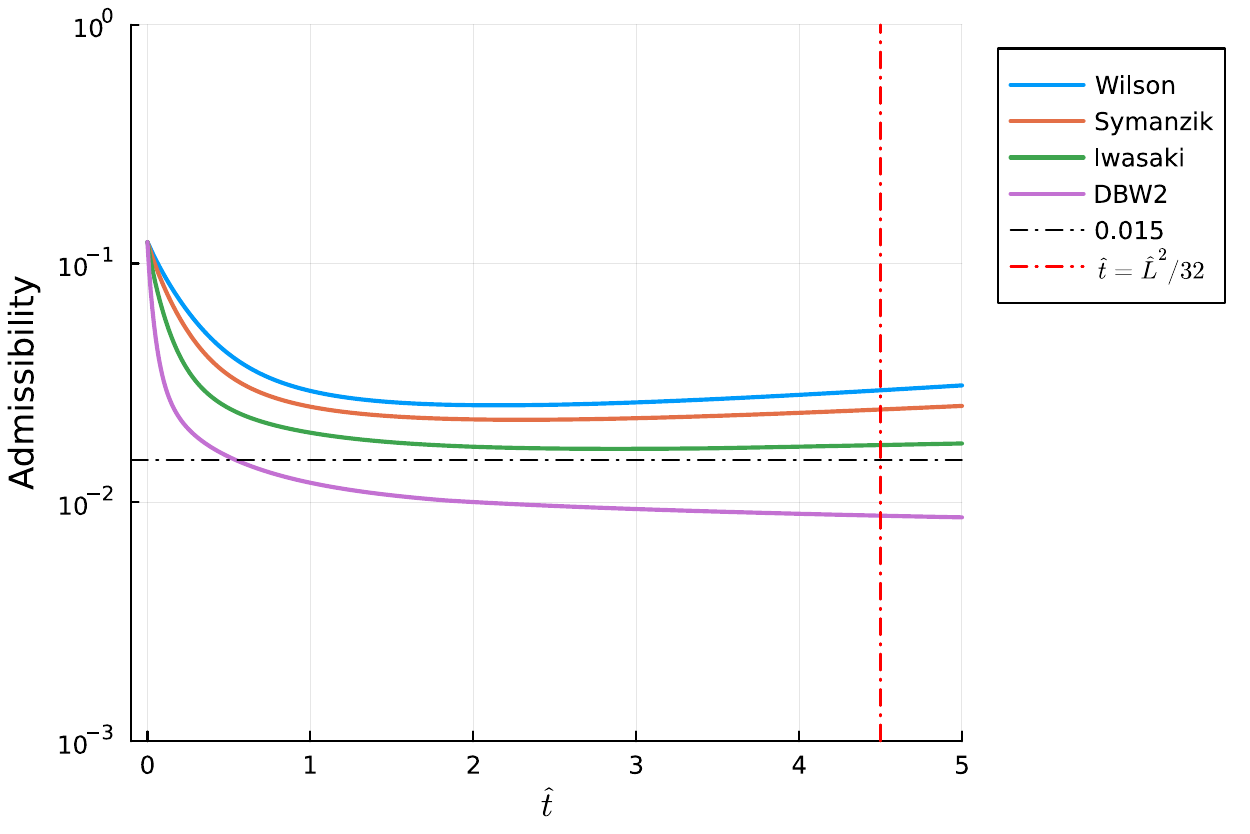}
\end{minipage}}
\fbox{\begin{minipage}[b]{0.48\columnwidth}
    \includegraphics[width=1.0\columnwidth]{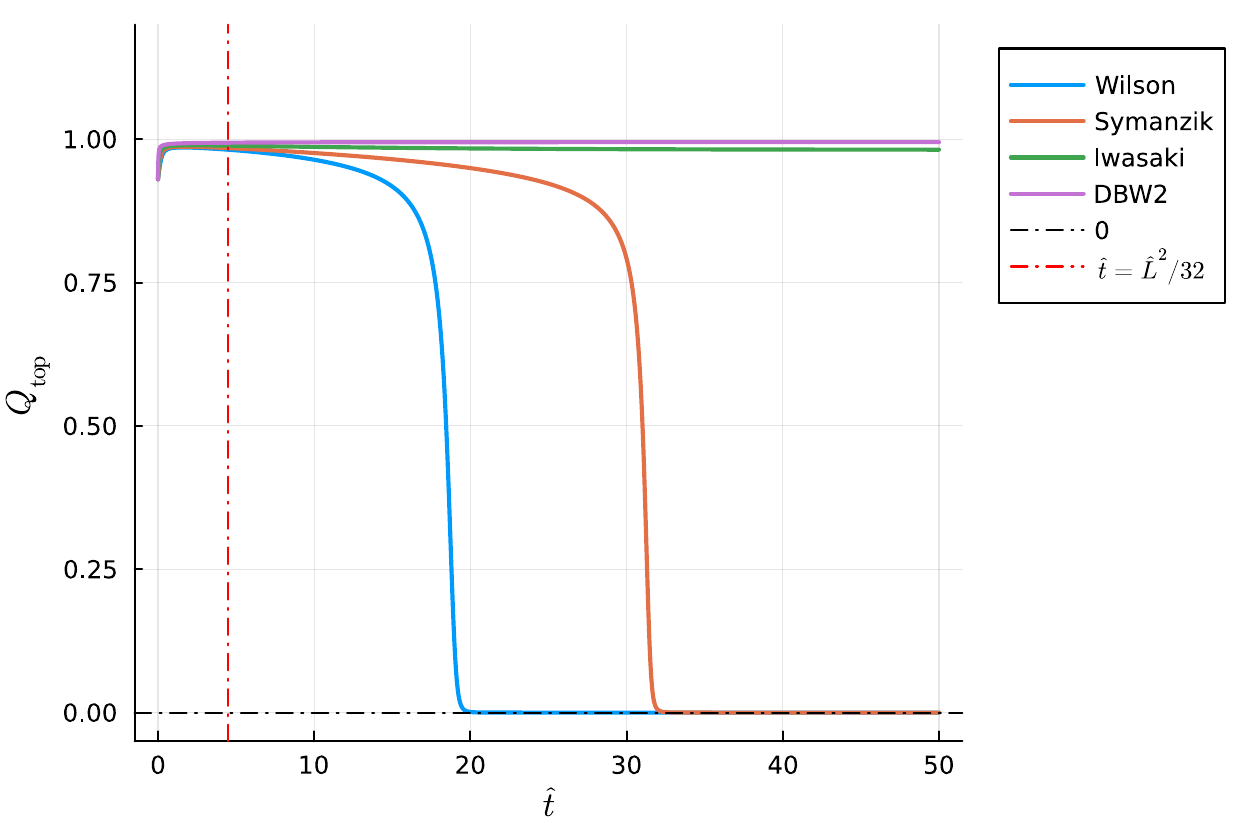}\\
    \includegraphics[width=1.0\columnwidth]{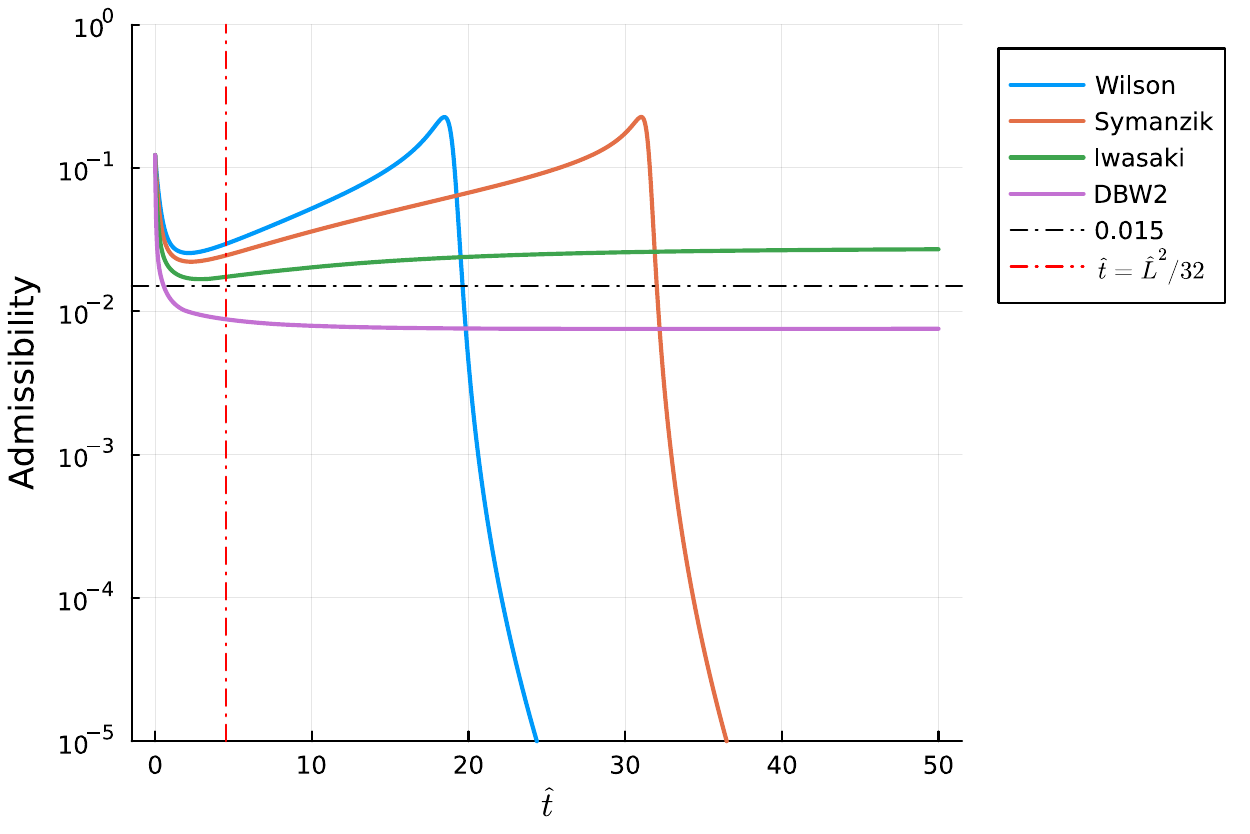}
\end{minipage}}
\caption{
Topological charge and the admissibility under flows for an instanton with $R = 3.0$.
The horizontal axis is dimensionless flow time and the vertical axis is the value of $Q_{\rm top}$ or $P_{\mathrm{adm}}$.
The left panel shows the short-time flow, $0\le \hat{t}\le 5$. The right panel shows the long-time flow, $0\le \hat{t}\le 50$.
\label{fig:flow_with_instanton_R=3.0}
}
\end{figure}

Figure~\ref{fig:flow_with_instanton_R=3.0} shows the gradient flows for the one-instanton configuration with $R=3.0$.
This instanton is large enough, and all the flows, Wilson, Symanzik, Iwasaki, and DBW2, detect the topological charge $Q_{\mathrm{top}}\approx 1$ in the short-time range, $0.5\lesssim \hat{t} \le 5$, as we can see in the left panel. For large instantons, short-time behaviors look quite similar among these flows at the qualitative level. 

The right panel of Figure~\ref{fig:flow_with_instanton_R=3.0} shows the long-time behavior of the flow, $0<\hat{t}<50$, for $R=3.0$, and we can see the qualitative difference between $1+12 c_1< 0$ and $1+12 c_1\ge 0$. 
For Iwasaki and DBW2 ($1+12 c_1<0$), the topological charge stays almost constant, $Q_{\mathrm{top}}\approx 1$, and the one-instanton configuration is maintained. 
This fits the expectation based on the classical continuum analysis in Section~\ref{sec:AnalyticalStudy_StabilityInstanton}. 
For the DBW2 flow, the admissibility measure $P_{\mathrm{adm}}$ becomes less than L\"uscher's bound $0.015$ around $\hat{t}\gtrsim 0.5$, and it stays almost constant around $P_{\mathrm{adm}}\lesssim 0.01$. 
For the Iwasaki flow, $P_{\mathrm{adm}}$ decreases and takes the minimal value slightly above L\"uscher's bound around $\hat{t}\approx 3$. After $\hat{t}\gtrsim 5$, $P_{\mathrm{adm}}$ stays around $0.015$--$0.03$ with showing the very mild growth.  

Next, let us look at the Wilson and Symanzik flows that have $1+12 c_1\le 0$. As predicted in Section~\ref{sec:AnalyticalStudy_StabilityInstanton}, $Q_{\mathrm{top}}$ shows the jump from $1$ to $0$ and the instanton disappears. 
For both Wilson and Symanzik flows, the admissibility measure $P_{\mathrm{adm}}$ first becomes smaller and takes the local minimum around $\hat{t}=2$, and it starts to grow gradually. 
$P_{\mathrm{adm}}$ takes the peak value around $P_{\mathrm{adm}}\approx 0.22$ when $Q_{\mathrm{top}}$ rapidly changes from $1$ to $0$. 
For the Wilson flow, this jump occurs around $\hat{t}\approx 18.5$, and for the Symanzik flow, the jump happens around $\hat{t}\approx 31$. 

\begin{figure}[t]
\begin{center}
\includegraphics[width=0.6\textwidth]{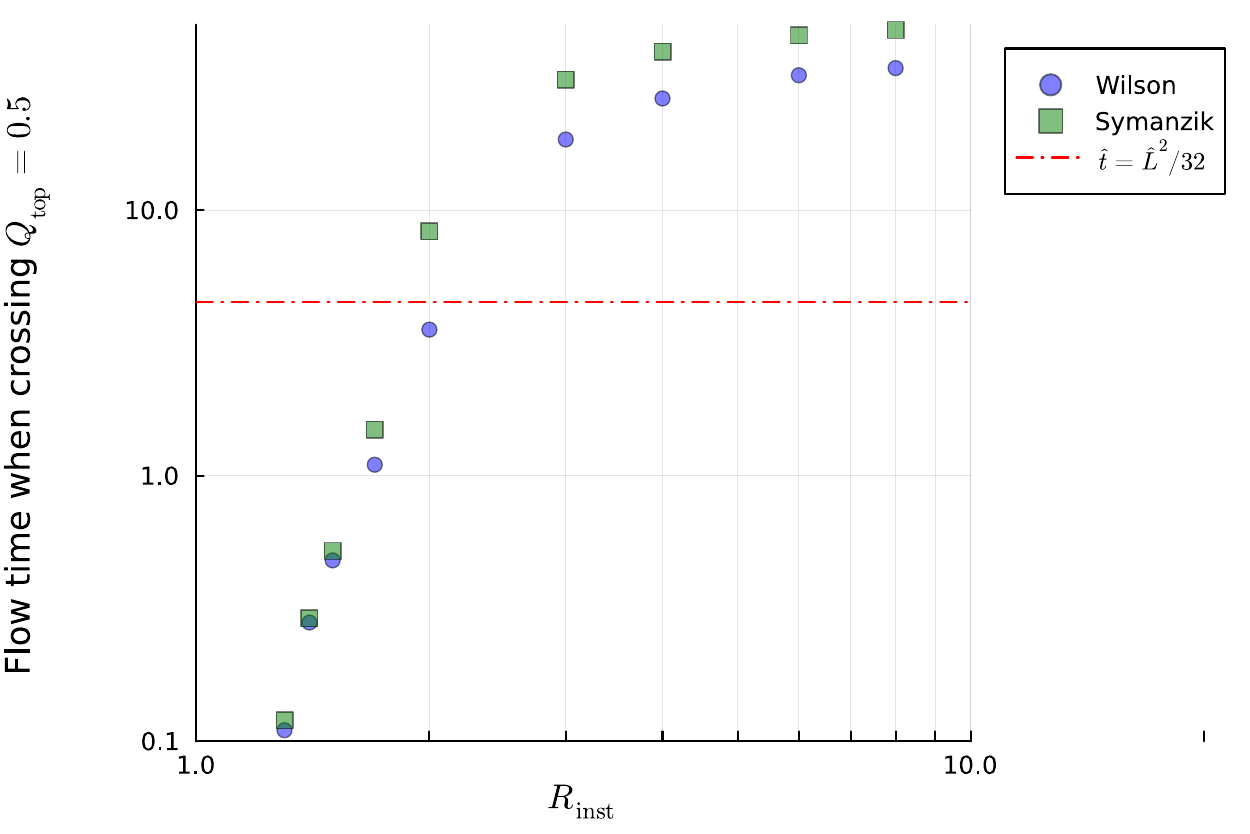}
\end{center}
\caption{
The lifetime of an instanton along with the flow time for Wilson and Symanzik flow.
The vertical axis is the flow time when crossing $Q_\mathrm{top} = 0.5$.
The horizontal axis is the instanton radius.
Large instanton lives longer.
}
\label{fig:Qshrink}
\end{figure}

In Figure~\ref{fig:Qshrink}, we show the $R$-dependence for the flow time $\hat{t}_*$ at which the instanton shrinks and slips away for the Wilson and Symanzik flows. We determine $t_*$ by the time $Q_{\mathrm{top}}$ crosses $0.5$. 
This figure shows that large instantons also disappear in a finite time $\hat{t}$ for Wilson and Symanzik flows. 
If $L=\infty$, $t_*$ is expected to scale as $R^4$ for the Wilson flow and as $R^6$ for the Symanzik flow, but they show the saturation due to the finite volume effect for $L=12$.

\subsection{Flows on gauge configurations with \texorpdfstring{$SU(2)$}{SU(2)} pure Yang-Mills theory}
\label{sec:Numerics_MonteCarlo}

Here, we perform simulation for $SU(2)$ pure Yang--Mills theory with the Wilson action for $\beta \,(=4/g^2) = 2.45$ in $L^4=12^4$.
This corresponds to $a\simeq (2.1 \,\text{GeV})^{-1}\simeq 0.093 \,\text{fm}$ and the physical lattice size is $L a \simeq 1.1$ fm \cite{Hirakida:2018uoy}.
Gauge configurations are generated using the heatbath algorithm with the plaquette gauge action. After each heatbath update, overrelaxation is applied three times. We have performed 50,000 updates, and observables are measured for every 20 configurations or more. The autocorrelation time for the unsmeared topological charge is approximately 1.4 (see Appendix \ref{App:generation}). 
The number of analyzed configurations is 200.
Configurations are generated with  \texttt{LatticeQCD.jl} in JuliaQCD \cite{Nagai:2024yaf}.

\subsubsection{Short-time and long-time behaviors of gradient flows}

\begin{figure}[t]
\centering
\includegraphics[width=0.75\textwidth]{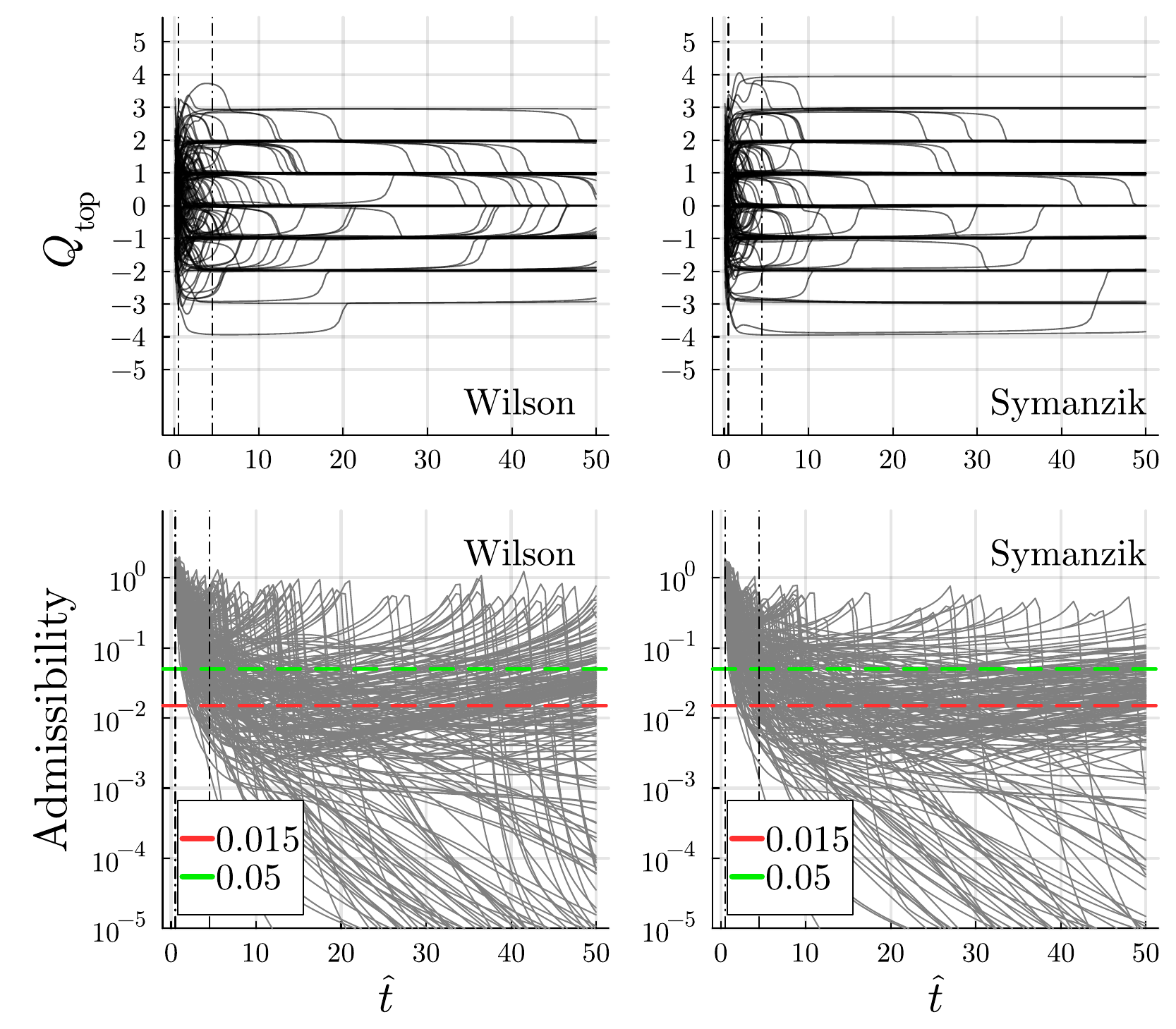}
\caption{
Flow time dependence of $Q_{\rm top}$ ({\it upper row}) and admissibility ({\it lower row}) for Wilson and Symanzik flows.
The vertical dash-dotted lines represent the conventional bound at $\hat{t} = 0.5$, $4.5$ in \eqref{eq:conventional_bound}.
The horizontal dashed lines for figures on the lower row are guidelines, including the estimate of admissibility bound by L\"uscher~\cite{Luscher:1981zq}.
}
\label{fig:Q&Adm_flowtime_dep_1}
\end{figure}

\begin{figure}[th]
\centering
\includegraphics[width=0.75\textwidth]{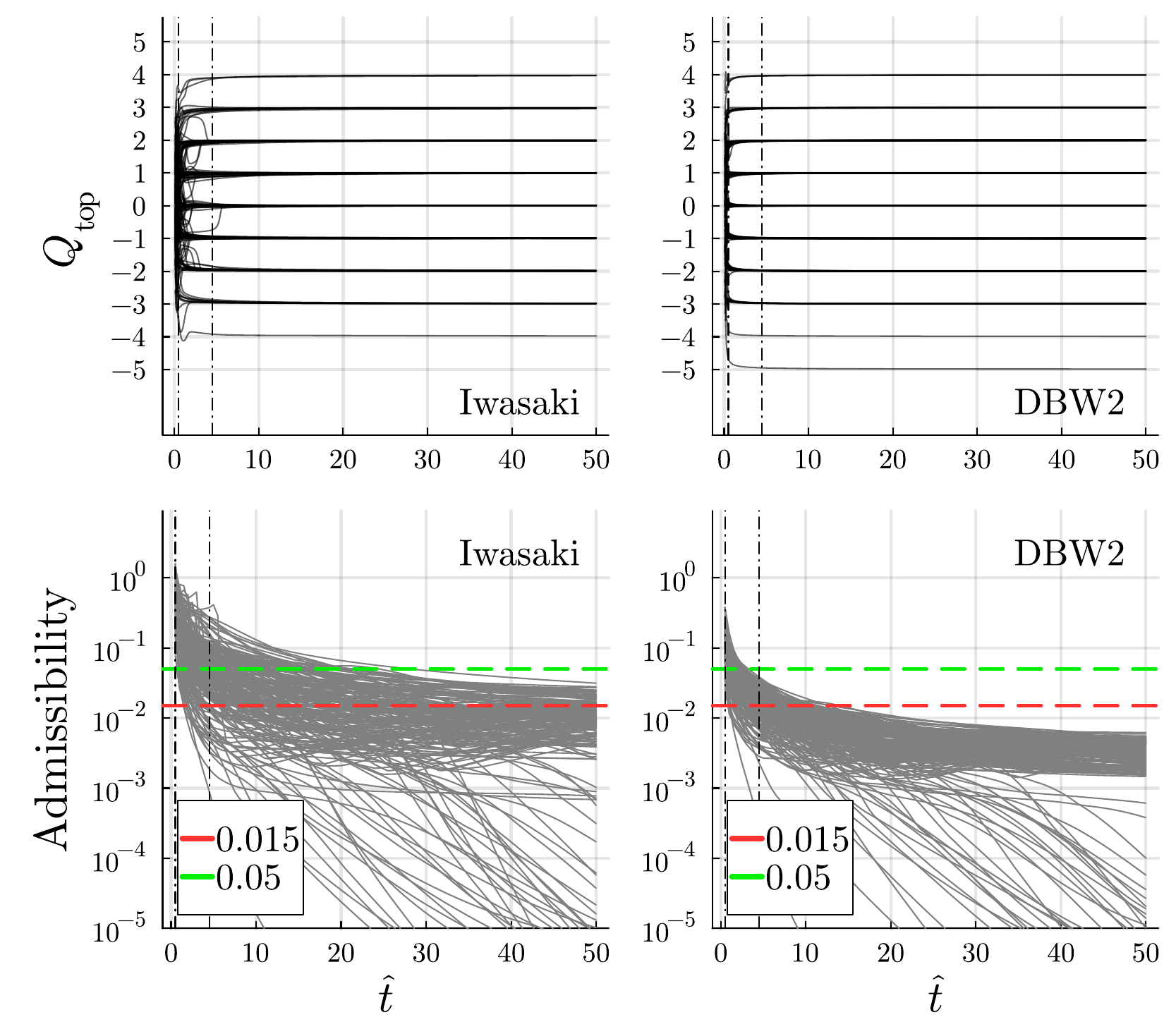}
\caption{
Flow time dependence of $Q_{\rm top}$ ({\it upper row}) and admissibility ({\it lower row}) for Iwasaki and DBW2 flows.
The vertical dash-dotted lines represent the conventional bound at $\hat{t} = 0.5$, $4.5$ in \eqref{eq:conventional_bound}.
The horizontal dashed lines for figures on the lower row are guidelines, including the estimate of admissibility bound by L\"uscher~\cite{Luscher:1981zq}.
}
\label{fig:Q&Adm_flowtime_dep_2}
\end{figure}

\begin{figure}[th]
\centering
\includegraphics[width=0.7\textwidth]{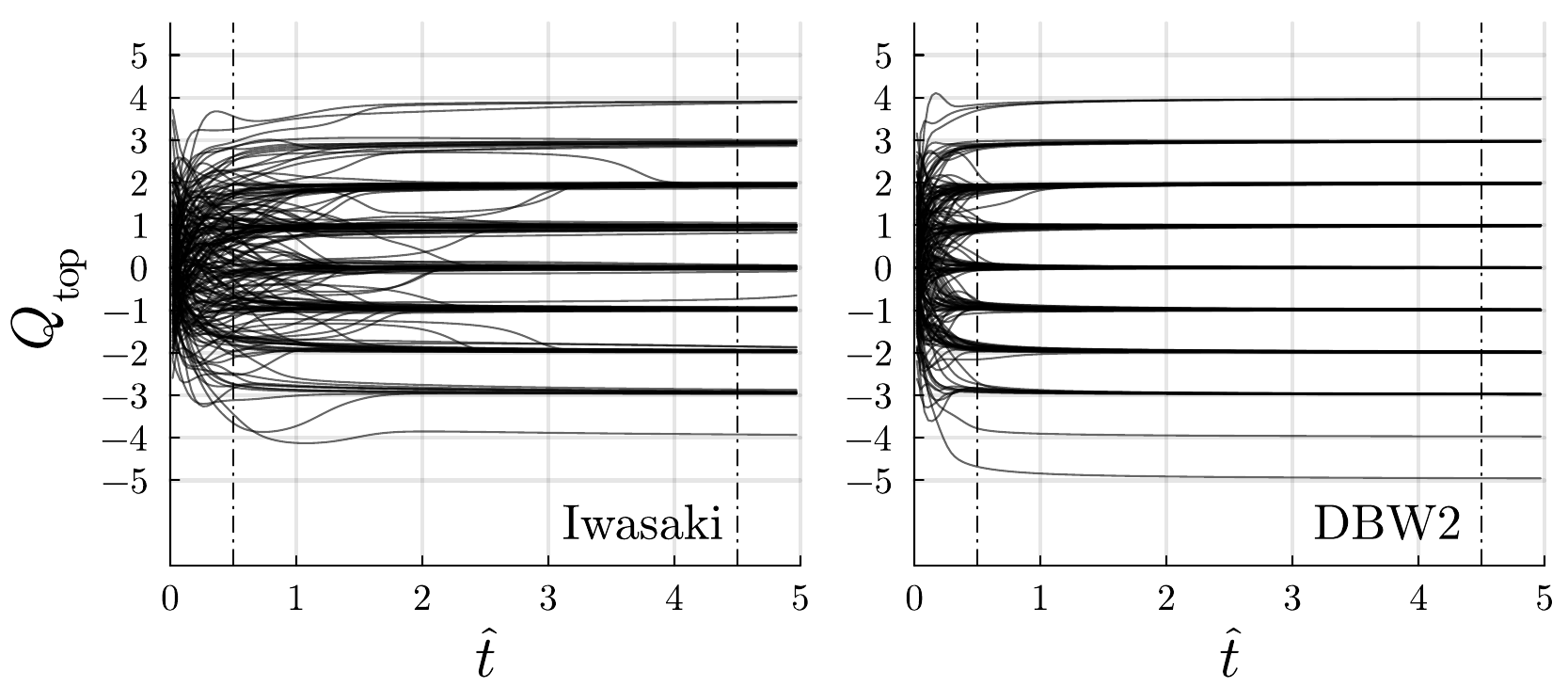}
\caption{
Zoomed plot of Figure~\ref{fig:Q&Adm_flowtime_dep_2} for $Q_\mathrm{top}$ on flow time $\hat{t} \le 5$. We can observe the fast convergence to the quantized topological charges for these flows. 
}
\label{fig:Q&Adm_flowtime_dep_3}
\end{figure}

Let us first show the smearing history of the topological charge of gauge configurations. 
Figure~\ref{fig:Q&Adm_flowtime_dep_1} shows the results of the flow time dependence of the topological charge $Q_{\mathrm{top}}$ and the admissiblity measure $P_{\mathrm{adm}}$ for Wilson and Symanzik flows. 
The dimensionless flow time is taken up to $\hat{t}=50$, and we show the conventional bound~\eqref{eq:conventional_bound}, $\hat{t}=0.5$ and $\hat{t}=4.5$, by vertical dash-dotted lines. 
For the admissibility measure $P_{\mathrm{adm}}$, we draw the horizontal line corresponding to L\"uscher's bound $P_{\mathrm{adm}}=0.015$ and also another line $0.05$ as a guideline. 

In both Wilson and Symanzik flows, the jump of topological sectors frequently occurs for any time range of $0<\hat{t}<50$. By comparing the figures for $Q_{\mathrm{top}}$ and $P_{\mathrm{adm}}$, one can see that $P_{\mathrm{adm}}$ has large peaks $\gtrsim 0.5$ when the topological charge rapidly changes. 
It seems that there is no clear way to determine when we should stop the gradient flows for the topological charge in the case of Wilson and Symanzik flows: In any range of the flow time within $\hat{t}<50$, we encounter some jumps of topological sectors. 

In Figures~\ref{fig:Q&Adm_flowtime_dep_2} and \ref{fig:Q&Adm_flowtime_dep_3}, we show the smearing history for Iwasaki and DBW2 flows, and the qualitative difference compared with Figure~\ref{fig:Q&Adm_flowtime_dep_1} (Wilson and Symanzik flows) can be clearly observed. 
Let us first look at the behaviors of the Iwasaki flow shown in the left panels. The topological charges start to localize around integers after $\hat{t}\gtrsim 2$, and there are several jumps of topological sectors until $\hat{t}\approx 6$, but its occurrence is much less frequent compared with the cases of Wilson and Symanzik flows. After $\hat{t}\gtrsim 6$, there is literally no jump of topological sectors.

The DBW2 flow has a more striking property as shown in the right panel of Figure~\ref{fig:Q&Adm_flowtime_dep_2} and Figure~\ref{fig:Q&Adm_flowtime_dep_3}. 
After $\hat{t}\gtrsim 1$, the topological sectors are quite stable, and we do not see their jumps at all. 
Therefore, Iwasaki and DBW2 flows stabilize not only the one-instanton configuration but also the topological sectors themselves within our numerical setup. 

\begin{figure}[t]
\centering
\includegraphics[scale=0.21]{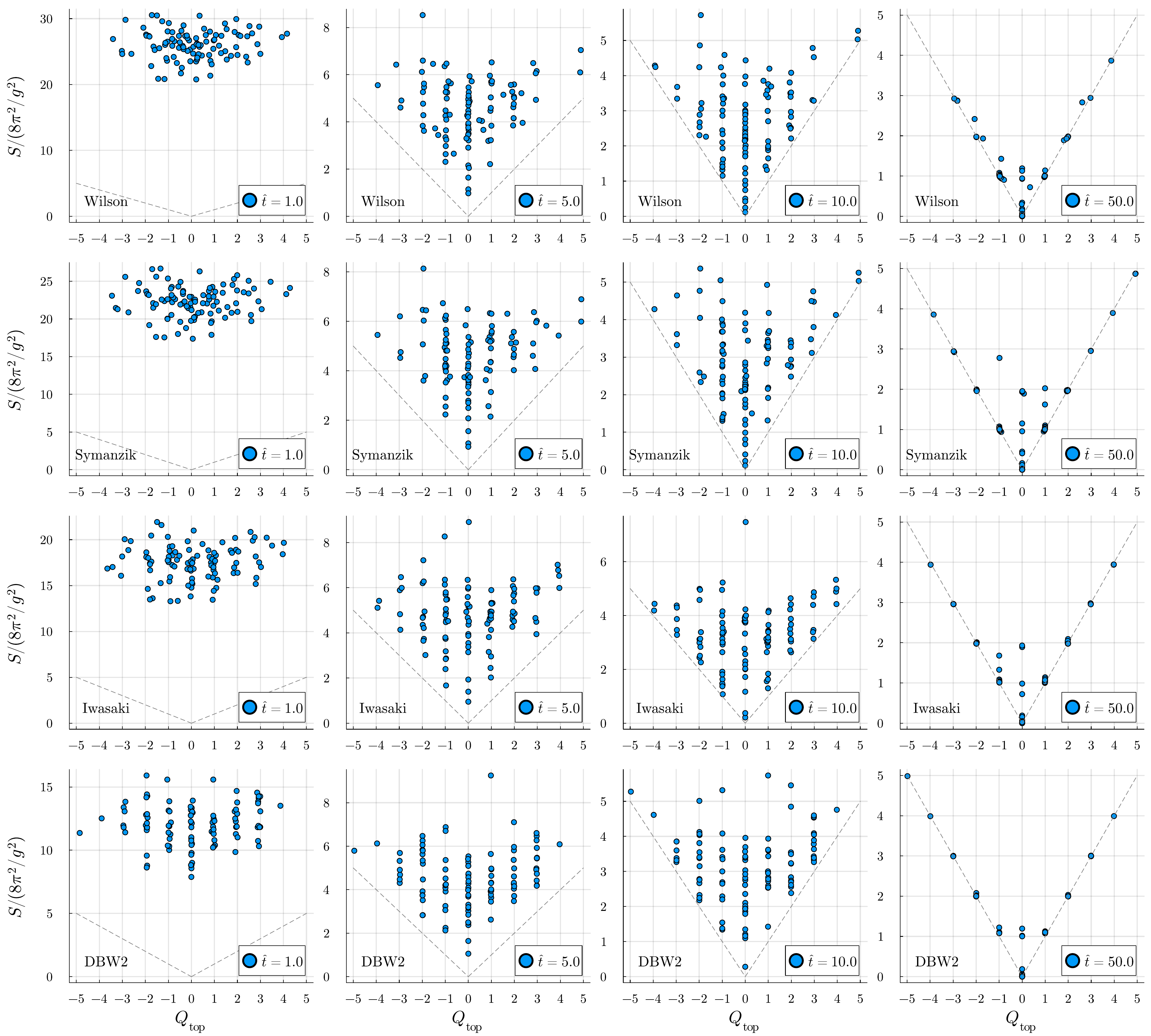}
\caption{Plots of $Q_{\rm top}$ versus $S/(8\pi^2/g^2)$ for each flow at flow time $\hat{t} = 1.0$, $5.0$, $10.0$, $50.0$.
The figures are drawn by $100$ data points, and the dashed lines show the BPS bound.
\label{fig:Bogomol'nyi bound}}
\end{figure}

Figure~\ref{fig:Bogomol'nyi bound} shows the scatter plot for the classical action versus the topological charge at the flow time $\hat{t}=1, 5, 10, 50$.~\footnote{A similar plot can be found in \cite{Kitano:2020mfk}.}
For the action, we evaluate the Wilson action and divide it by $8\pi^2/g^2$. The classical Yang-Mills action has the BPS bound, 
\begin{equation}
    S \ge \frac{8\pi^2}{g^2} |Q_{\mathrm{top}}|. 
\end{equation}
Thus, all the points in the scattered plots appear in the range $S/(8\pi^2/g^2)\ge |Q_{\mathrm{top}}|$ up to the lattice artifact. The equality holds if and only if the (anti-)self-dual Yang-Mills equation is satisfied. 

Comparing the plots for $\hat{t}=1$ (leftmost column of Figure~\ref{fig:Bogomol'nyi bound}), we can observe that the DBW2 flow already captures the topological sectors very clearly, and the Iwasaki flow also has its tendency while it is less sharp compared with DBW2. Wilson and Symanzik flows need more time to observe topological sectors. We also note that the magnitude of the action has a huge difference among these flows: For Wilson flows, $S/(8\pi^2/g^2)$ scatters around $20$--$30$, while for DBW2, it scatters around $8$--$17$ at $\hat{t}=1$. The value of $c_1$ strongly affects the short-time behavior of gradient flows. 

When the flow time becomes $\hat{t}=5$ (second left column of Figure~\ref{fig:Bogomol'nyi bound}), all the flows show qualitatively similar behaviors: The gauge configurations are mostly separated to topological sectors, and the value of action accumulates in the region $\lesssim 7$ in the unit of $8\pi^2/g^2$.  
When we look at them more closely, we can see that there are many configurations with non-integer topological charges for the Wilson flow at $\hat{t}=5$, and they are identified as the configuration on the way to jump between topological sectors. Those jumps also occur for Symanzik and Iwasaki flows, but they become less and less frequent as we change the flow from Wilson to Symanzik and from Symanzik to Iwasaki. 
For the Iwasaki flow, the jump of topological sectors is not observed for $\hat{t}=10$ in our configuration set, but it happens for Wilson and Symanzik flows even at $\hat{t}=10$. 

By continuing the flow until $\hat{t}=50$ (rightmost panel of Figure~\ref{fig:Bogomol'nyi bound}), many configurations saturate the BPS bound. Since Iwasaki and DBW2 do not have the jump of topological sectors at all, they may provide a stable numerical technique to find BPS solutions in various topological sectors.

\subsubsection{Topological susceptibilities with gradient flows}

\begin{figure}[t]
\begin{center}
\includegraphics[width=1.0\textwidth]{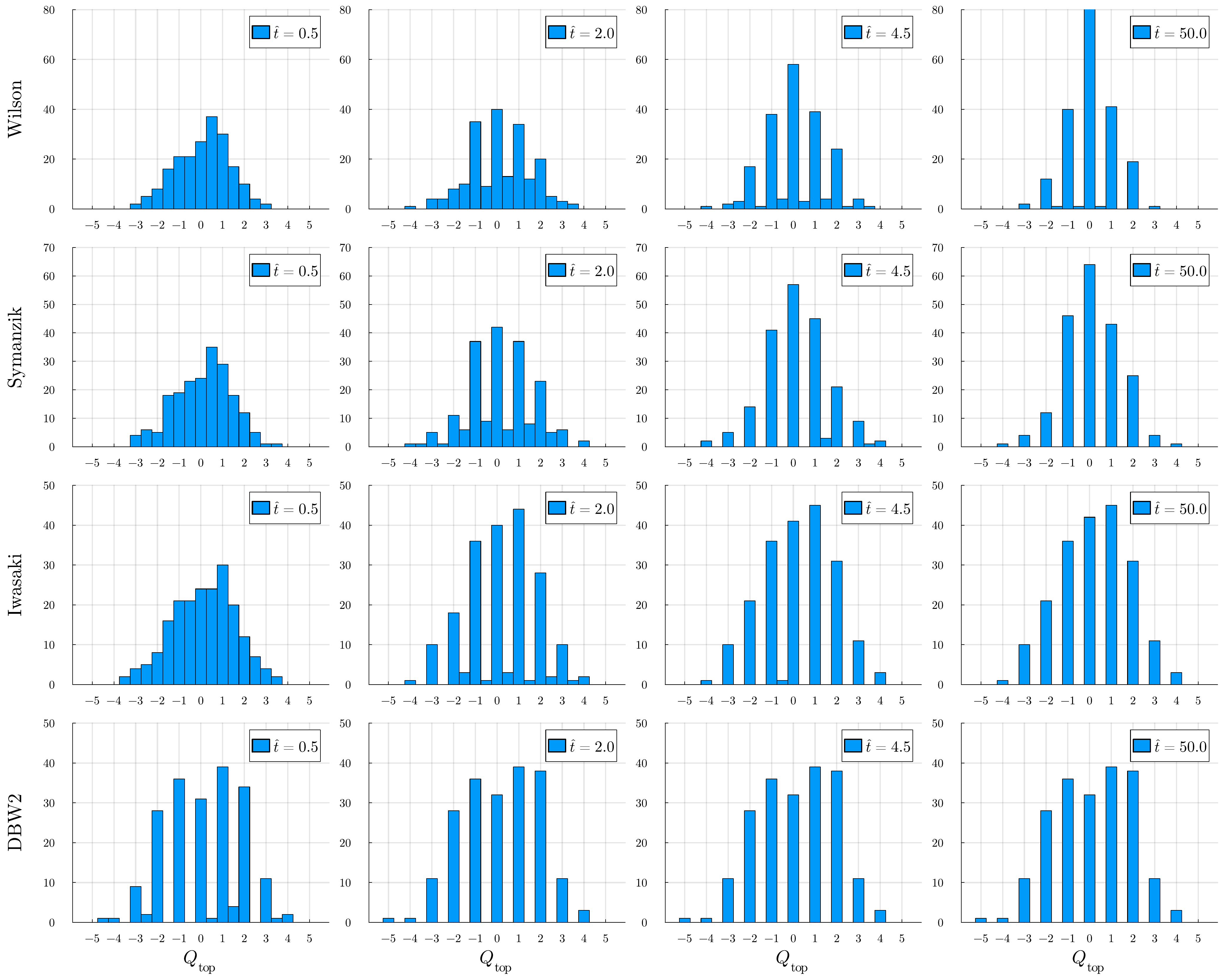}
\end{center}
\caption{
Histogram of the topological charge under flows.
From top to bottom, results for Wilson flow, Symanzik flow, Iwasaki flow, and DBW2 flow are shown.
From left to right, $\hat{t}=0.5$,
$\hat{t}=2.0$, $\hat{t}=4.5$,
$\hat{t}=50$ are shown. 
The asymmetry of the charge distribution would be due to the smallness of the ensembles. 
\label{fig:histograms}}
\end{figure}

Figure~\ref{fig:histograms} represents the histogram of the topological charge obtained by the flows at $\hat{t}=0.5, 2, 4.5$, and $50$. 
The histogram is roughly symmetric about $Q_{\mathrm{top}}=0$, and its violation would be due to the smallness of the number of configurations. 
As we have emphasized repeatedly, DBW2 (bottom panel) shows the quantization of topological charge at $\hat{t}=0.5$, and its distribution does not change little along the flow for $2\le \hat{t}\le 50$, which can be clearly confirmed by this figure for the histogram.
The Iwasaki flow also maintains the distribution of topological charge after it is quantized around $\hat{t}\approx 2$. 

For Wilson and Symanzik flows, we can observe that the distribution becomes more and more shrinking along the flow. Comparing the histogram between $\hat{t}=2$ and $4.5$ and between $\hat{t}=4.5$ and $50$, the number of configurations with smaller $Q_{\mathrm{top}}$ becomes larger through the process that $Q_{\rm top}$ ``spill out'' from a certain value, including integer, to the closest integer. 

\begin{figure}[t]
\begin{center}
\includegraphics[width=1.0\textwidth]{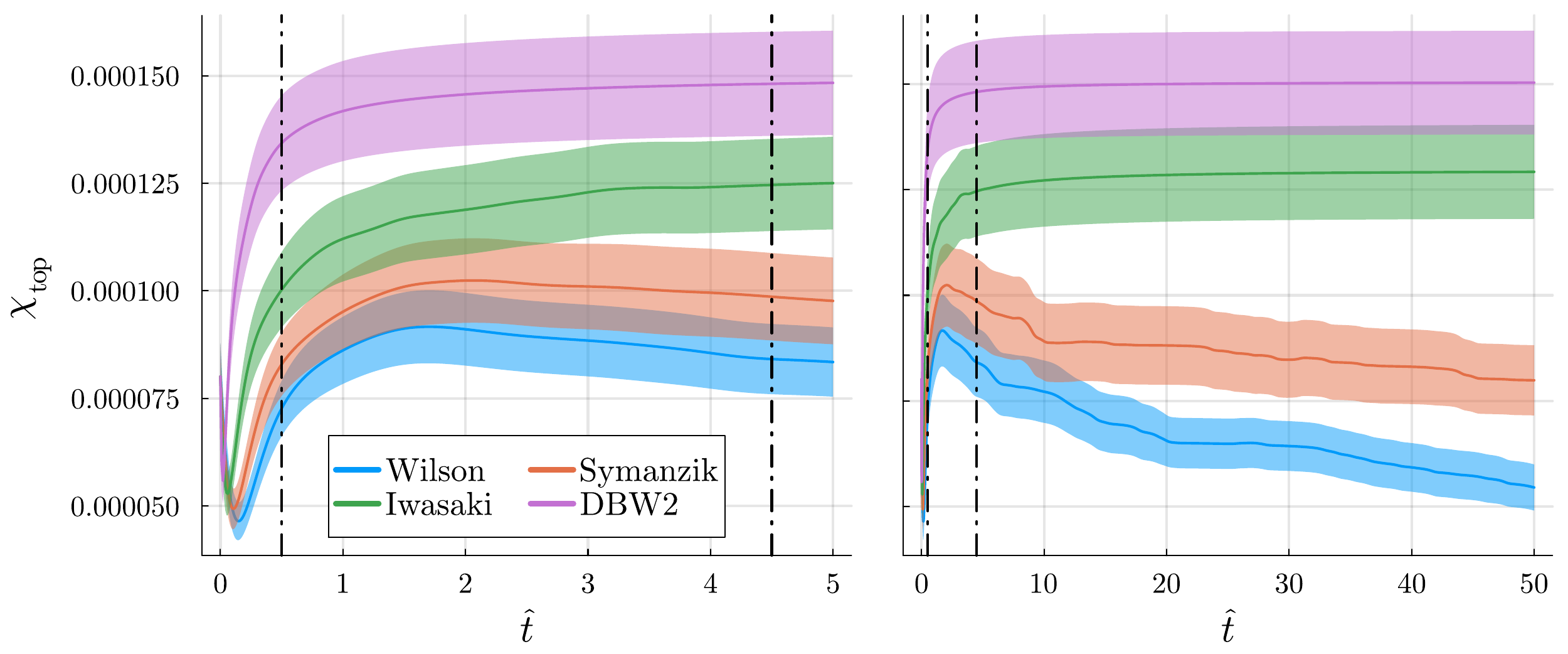}
\end{center}
\caption{
Flow time dependence of $\chi_{\rm top}$ for each flow.
Color bands represent statistical errors.
The vertical dash-dotted lines represent the conventional bound at $\hat{t} = 0.5$, $4.5$ in \eqref{eq:conventional_bound}.
{\it (Left)}: $\chi_{\rm top}$ for $\hat{t} = 0$ to $5$.
{\it (Right)}: $\chi_{\rm top}$ for $\hat{t} = 0$ to $50$.
\label{fig:flows_chi_top_along_t}}
\end{figure}

\begin{table}[tb]
\begin{center}
\begin{tabular}{c|cccccc}
Name & \( {\hat{t}}=0.5 \) & \( {\hat{t}}=2.0 \) & \( {\hat{t}}=4.5 \) & \( {\hat{t}}=10 \) & \( {\hat{t}}=30 \) & \( {\hat{t}}=50 \) \\ \hline\hline
Wilson   & $197(4)$ & $208(5)$ & $204(5)$ & $200(5)$ & $191(4)$ & $183(5)$ \\
Symanzik & $204(5)$ & $215(5)$ & $213(5)$ & $207(5)$ & $205(5)$ & $202(5)$ \\
Iwasaki  & $213(5)$ & $223(5)$ & $225(5)$ & $227(5)$ & $227(5)$ & $227(5)$ \\
DBW2     & $229(5)$ & $234(5)$ & $235(5)$ & $236(5)$ & $236(5)$ & $236(5)$ \\
\end{tabular}
\end{center}
\caption{
Summary of $(\chi_{\rm top})^{1/4}$ MeV under several flows in the physical unit.
A result for ${t}=0$ is $202(5)$ MeV.
\label{tab:dim_chi_top_results}
}
\end{table}

Using the topological charge distribution, we can compute the topological susceptibility $\chi_{\mathrm{top}}$. 
Let us show the results of the flow-time dependence of $\chi_{\mathrm{top}}$ for different flows in Table~\ref{tab:dim_chi_top_results} and Figure~\ref{fig:flows_chi_top_along_t}, as they would be useful to understand the tendency of each flow better than just looking at histograms.
We should note, however, that the results shown here are obtained with $200$ samples, which is not necessarily sufficiently large, due to the numerical cost of computing gradient flows for long flow times. 

Figure~\ref{fig:flows_chi_top_along_t} shows the topological susceptibility $\chi_{\mathrm{top}}$ along with the flow time $\hat{t}$.
In our numerical setup, we observe the strong flow-action dependence of $\chi_{\mathrm{top}}$, resulting in different values and distinct curve behaviors concerning flow time $\hat{t}$.
For the Wilson flow, a plateau-like structure appears between $\hat{t} = 1$ and $4$, though there is a peak around $\hat{t} = 1.7$, after which the values start to decrease.
The Symanzik flow yields values higher than those of the Wilson flow but exhibits a similar qualitative trend. The peak is located in a similar region to the Wilson flow, followed by a decreasing trend.
In the Iwasaki flow, $\chi_{\mathrm{top}}$ converges around $\hat{t} = 5$ to $10$, stabilizing at a constant value thereafter. This value is higher than that obtained with the Symanzik flow and shows no decreasing trend.
For the DBW2 flow, $\chi_{\mathrm{top}}$ converges quickly, around $\hat{t} = 1$ to $3$, making it the fastest to reach a stable value. While similar in trend to the Iwasaki flow, it produces a higher value than the Iwasaki flow and does not decrease. 
All these properties are consistent with what we have seen on the topological charge distributions. 

The dimensionful topological susceptibility $(\chi_{\mathrm{top}})^{1/4}$ MeV as a function of $\hat{t}$ is shown in Table~\ref{tab:dim_chi_top_results}. 
The errors are computed by the jackknife estimation with unity bin size.
To compare our results, let us mention the results of 
several previous studies for topological susceptibility in similar setups: 
For example, \cite{Kronfeld:1986ts} reported $(\chi_{\mathrm{top}})^{1/4} = 262(1) \ \mathrm{MeV}$ at $\beta = 2.4$. Studies using a fixed-point action, such as \cite{DeGrand:1996zb,DeGrand:1996mk}, have reported $(\chi_{\mathrm{top}})^{1/4} = 235(10) \ \mathrm{MeV}$, while \cite{deForcrand:1997esx} reported $(\chi_{\mathrm{top}})^{1/4} = 200(15) \ \mathrm{MeV}$. 
Since we have investigated the topological charge and its susceptibility at the region far from the continuum regime, it is not easy to give a fair statement for the above comparison. 
Moving on to the large volumes and weak coupling is therefore desired at which the notion of topological sector works properly.

To avoid potential confusion, we would like to emphasize that it is an ill-defined question to ask which flows give the ``correct'' topological susceptibility before taking the continuum limit. The results of different gradient flows should converge to the same value only after taking the continuum limit, and the difference caused by the choice of gradient flows should be understood as a part of the lattice artifacts for a given lattice spacing. 
It is important to understand how the topological susceptibilities obtained by different flows converge to the same physical values, and it should be interesting to see whether our observation of the topological stability and the fast convergence for the Iwasaki and DBW2 flows provide any practical efficiency. We leave it for future study to investigate that possibility more systematically. 

\section{Summary and discussion}
\label{sec:Summary}

In this paper, we have discussed the effect of the different flow actions on the topological charge distributions. 
By considering the classical continuum limit, we argue that the leading lattice correction by the dimension-$6$ operator needs to have a positive coefficient to stabilize the instantons at finite lattice spacing. 
After confirming this analytic prediction, we tested the stability of topological sectors for the actual gauge configurations for the $SU(2)$ Wilson gauge theory. 
We use the four distinct flow actions, Wilson, Symanzik, Iwasaki, and DBW2, and study the behaviors of the topological charge under these flows with monitoring the admissibility measure. 
We find that the Iwasaki and DBW2 flows (with $1+12 c_1<0$) make the topological sectors quite stable in the long flow times, while the Wilson and Symanzik flows (with $1+12 c_1\ge 0$) continuously experience the jump between topological sectors. 
As a byproduct of this study, we constructed numerical BPS solutions for various topological sectors. 

We emphasize that the Iwasaki and DBW2 flows not only stabilize topological sectors but also show fast smearing. In particular, the DBW2 flow separates the topological sectors clearly at the flow time $\hat{t}\gtrsim 0.5$, which almost corresponds to the lower side of the conventional bound~\eqref{eq:conventional_bound}. 
Thanks to this feature, the gradient flow with the DBW2 (and Iwasaki) actions may provide a numerically efficient way to study the topological property of $4$d gauge theories. 

Let us discuss several future directions of this study. Our numerical computation is performed for the $SU(2)$ Yang-Mills theory with the gauge coupling $\beta=2.45$ and the box size $L^4=12^4$. With this box size, the conventional bound \eqref{eq:conventional_bound} becomes $0.5\ll \hat{t}\ll L^2/32=4.5$, and the lower and upper bounds are not sufficiently separated. 
It is important to investigate the difference of gradient flows in larger boxes and to understand whether Iwasaki and DBW2 flows show a stable range of topological charge distribution within the conventional bound. Additionally, exploring the correlation between the index of the overlap Dirac operator and the topological charge~\cite{Chiu:2019xry} with flows would be another interesting avenue for further study.

It is very curious to know the behaviors for larger $\beta$, or weak lattice couplings $g^2$. As the expectation value of the plaquette action behaves as $\langle \tr(\bm{1}-U_{x,\mu\nu})\rangle \approx \frac{3}{8}g^2$, the local color-electromagnetic fluctuations are more suppressed as we take larger $\beta$. 
This implies the typical size of instantons must become larger compared with the lattice spacings. As a result, even for Wilson and Symanzik flows, it takes more time to shrink those topological objects, and the jump between topological sectors becomes less frequent \cite{Berg:2017tqu}.
It is important to see concretely how the different gradient flows converge to the identical topological charge distribution as $\beta$ is increased, and this knowledge will enable us to estimate and control systematic error for the topological susceptibility due to the choice of the gradient flow. 

It is worth mentioning the relation between this study and the empirical observation of topology freezing in Monte Carlo simulations. 
When performing Hybrid Monte Carlo (HMC) with the DBW2 gauge action, one observes fewer topological tunneling events than other gauge actions like the Wilson plaquette gauge action~\cite{Aoki:2002vt,McGlynn:2014bxa}.
This phenomenon is often attributed to the DBW2 gauge action being closer to the continuum limit as a nature of the perfect action. Our observations in this study are consistent with this understanding because HMC involves molecular dynamics steps that follow the classical equations of motion (with momenta initialized randomly). This freezing property is usually considered a drawback for generating gauge configurations, but our study suggests it is advantageous for smoothing gauge fields.

Our results may have an impact on the numerical studies about the topological properties of QCD. 
Because of its importance, there is a huge endeavor for studying the topological susceptibility~\cite{Durr:2001ty, Chiu:2011dz, Trunin:2015yda, Dromard:2016psw, Frison:2016vuc, Aoki:2017paw, Alexandrou:2017bzk, Matsumoto:2019jia, DeGrand:2020utq, Lombardo:2020bvn, Chen:2022fid}. 
At finite temperatures, there is a theoretical proposal that anomalous $U(1)_A$ transformation becomes approximately good symmetry when the flavored chiral symmetry is restored~\cite{Shuryak:1993ee, Cohen:1996ng, Cohen:1997hz, Aoki:2012yj}, 
which possibility has been investigated for a long time in the lattice numerical simulations using chiral fermions~\cite{HotQCD:2012vvd, Cossu:2013uua, Buchoff:2013nra, Dick:2015twa, Taniguchi:2016tjc, Tomiya:2016jwr, Ding:2020xlj, Aoki:2020noz, Aoki:2021qws, Dini:2021hug, Gavai:2024mcj}. 
Many of these works smear the gauge fields using the Wilson and Symanzik flows (or other smearing methods, such as stout smearing that approximates the Wilson flow), but our observations imply that re-analysis using the Iwasaki and DBW2 flows may lead to different results due to their stability about topological sectors. 
In this regard, we should extend our study to the $SU(3)$ gauge group with including dynamical fermions. 

Lastly, we would like to note that the gradient flow can be used to implement the coarse-graining for the renormalization group~\cite{Luscher:2013vga, Makino:2018rys} and also provides a convenient way for the composite operator renormalization of lattice operators, such as the energy-momentum tensor~\cite{Suzuki:2013gza, Makino:2014taa} and the quasi-parton distribution function~\cite{Monahan:2016bvm}.
Although we focused on the topological property in this paper, it would be interesting to explore if our observations of the fast convergence of the Iwasaki and DBW2 flows shed new light on these aspects of the gradient flow.

\section*{Acknowledgments} 
This project was started when the authors enjoyed having pizzas in a restaurant near Kyoto University, and the authors sincerely thank Pizzeria da Ciro.
The authors appreciate useful discussions with Yasumichi Aoki, Issaku Kanamori, Masakiyo Kitazawa, and especially Okuto Morikawa. 
A.T.\ is grateful to the collaborators in \cite{Kagimura:2015via} and to Taku Izubuchi, whose discussions during this period provided valuable insights for this work.
The numerical calculations were carried out on XC40 at YITP in Kyoto University and Ajiro in Center
for Computational Sciences at the University of Tsukuba.
Our numerical codes are based on JuliaQCD \cite{Nagai:2024yaf}, and A. T. is grateful to Yuki Nagai for maintaining it.
Discussions during the long-term workshop, HHIQCD2024, at Yukawa Institute for Theoretical Physics (YITP-T-24-02) were helpful in completing this work.
The work of Y.T.\ was partially supported by Japan Society for the Promotion of Science (JSPS) KAKENHI Grant No. 23K22489, and by Center for Gravitational Physics and Quantum Information (CGPQI) at YITP.
The work of A.T.\ was partially supported by JSPS  KAKENHI Grant Numbers 20K14479, 22K03539, 22H05112, and 22H05111, and MEXT as ``Program for Promoting Researches on the Supercomputer Fugaku'' (Simulation for basic science: approaching the new quantum era; Grant Number JPMXP1020230411, and 
Search for physics beyond the standard model using large-scale lattice QCD simulation and development of AI technology toward next-generation lattice QCD; Grant Number JPMXP1020230409).

\appendix
\section{More on configuration generation and measurement}\label{App:generation}
Here we describe the production of gauge configurations.
We use $\beta = 2.45$ for $L=12$ lattice.
The heatbath update is 50000 times with 3 times overrelaxation for each heatbath update. 
Our lattice spacing is $a \approx 0.093$ fm, and this is slightly coarse.
This is good for measuring the topological observable because this system is apart from the critical regime.
Configurations are generated with \texttt{LatticeQCD.jl} in JuliaQCD \cite{Nagai:2024yaf}.

\begin{figure}[t]
\begin{center}
\includegraphics[scale=0.5]{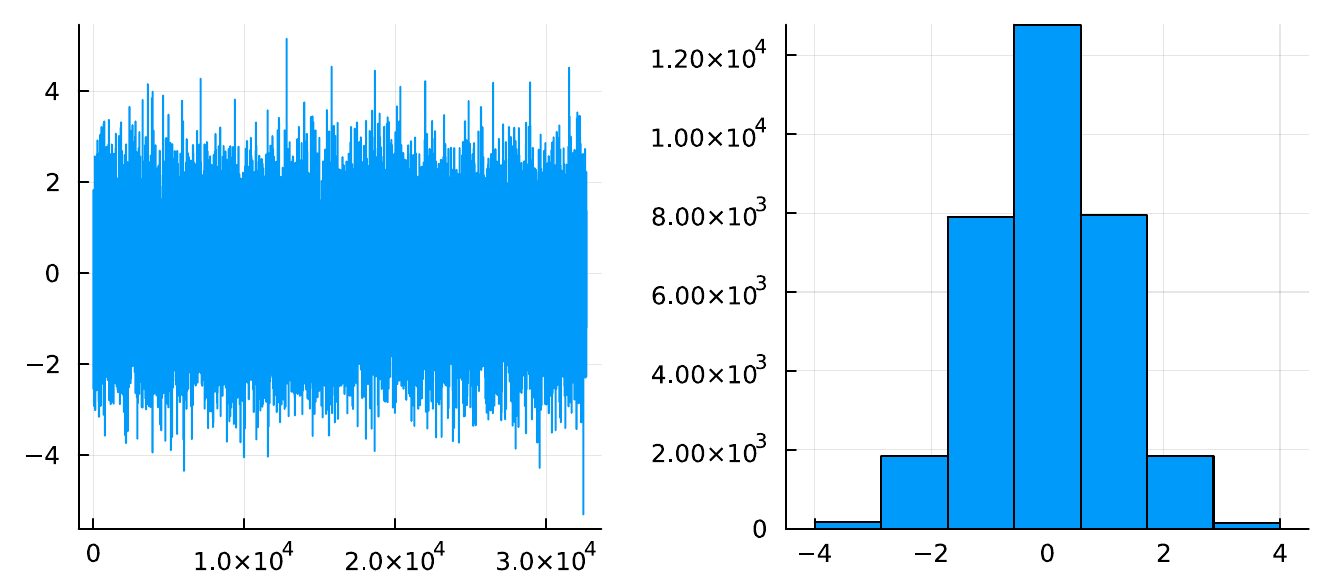}
\end{center}
\caption{Topological charge on the fly.
The topological charge is measured without flows.
{\it (Left)} History.
Topology fluctuates well.
{\it (Right)} Histogram.
The topological charge is well sampled.
\label{fig:top-hist}}
\end{figure}

\begin{figure}[t]
\begin{center}
\includegraphics[scale=0.5]{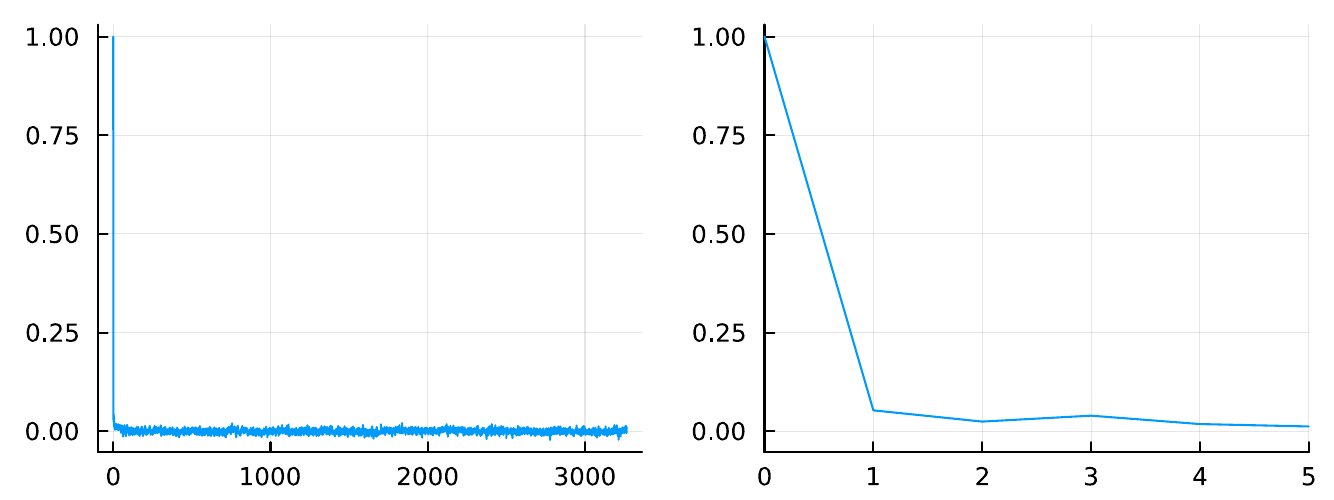}
\end{center}
\caption{Autocorrelation function for $Q_{\rm top}$ without flows, which is measured on the fly. 
The horizontal axis is the Monte-Carlo time $t_{\mathrm{MC}}$.
{\it (Left)} $t_\mathrm{MC}=0$ to $3100$.
{\it (Right)} $t_\mathrm{MC}=0$ to $5$.
\label{fig:acfunc}}
\end{figure}

\begin{figure}[t]
\begin{center}
\includegraphics[width=0.425\textwidth]{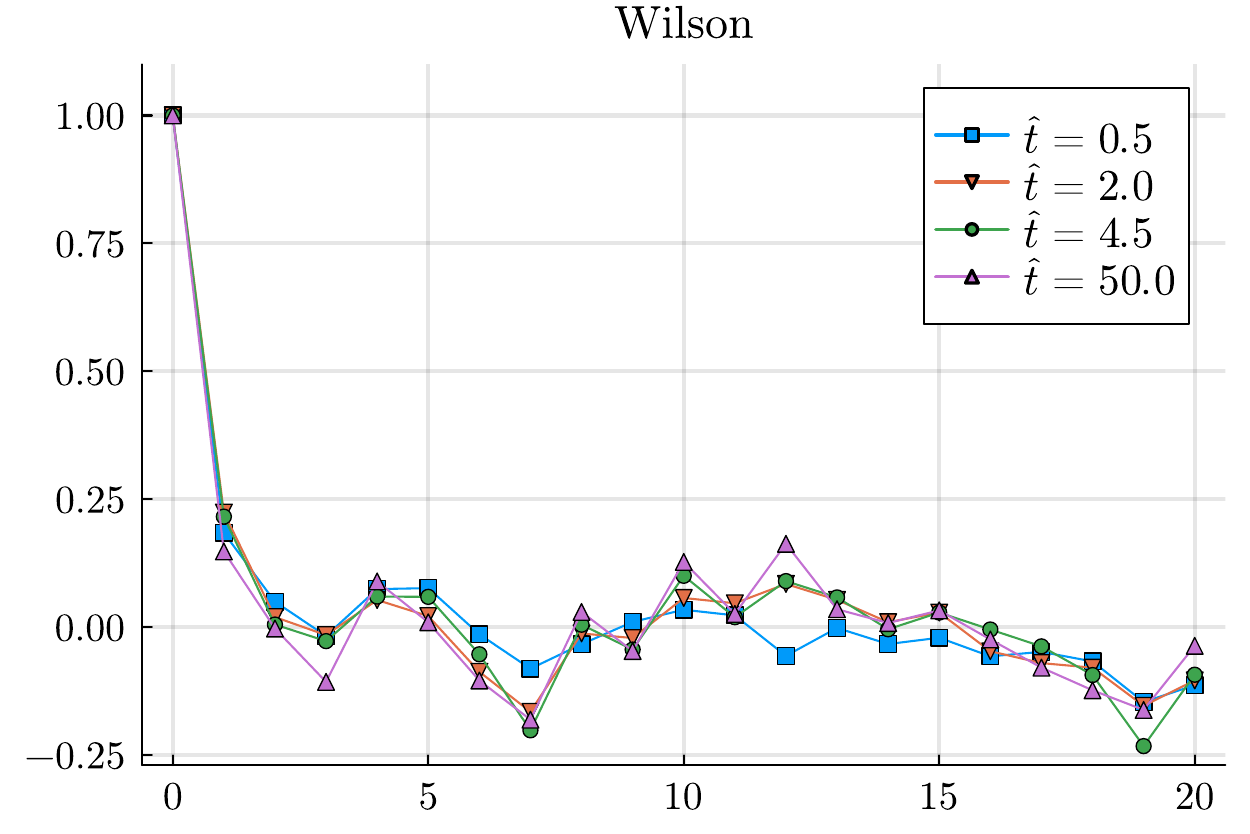}
\includegraphics[width=0.425\textwidth]{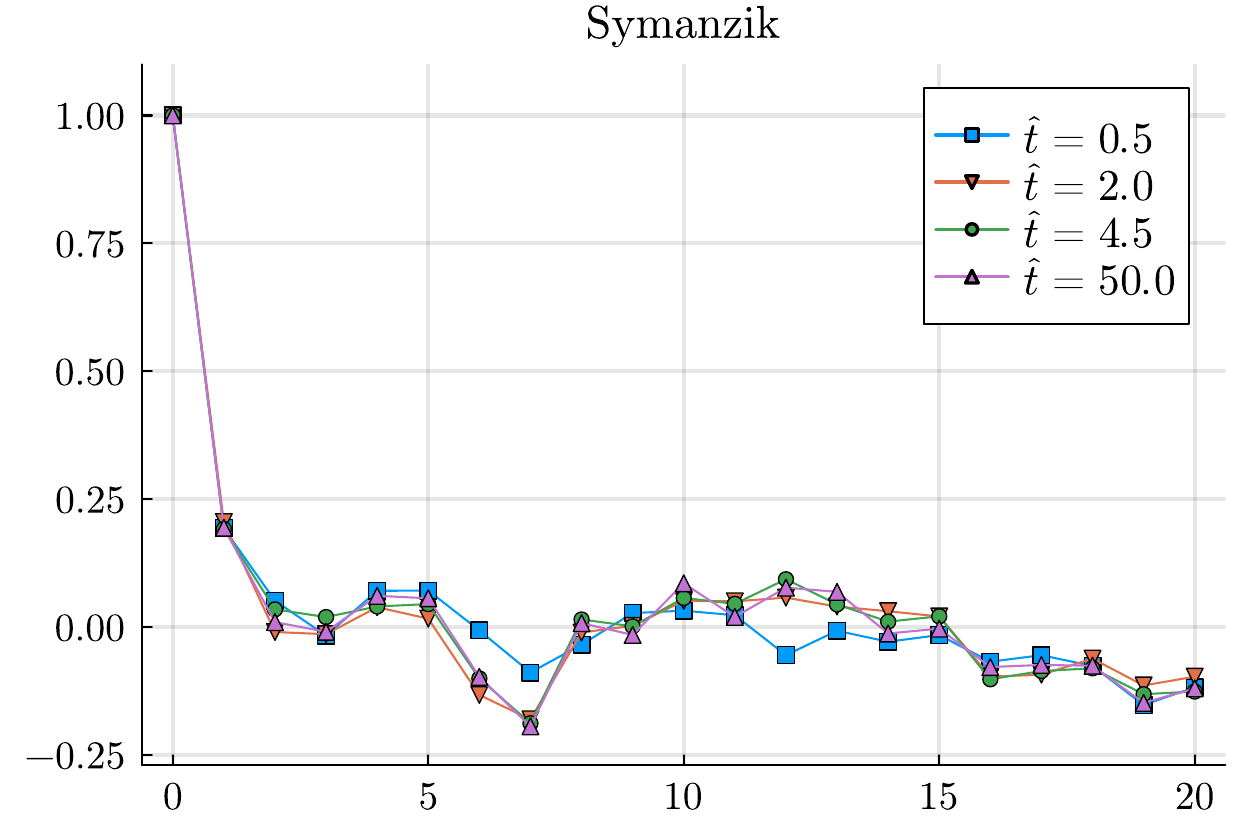}
\includegraphics[width=0.425\textwidth]{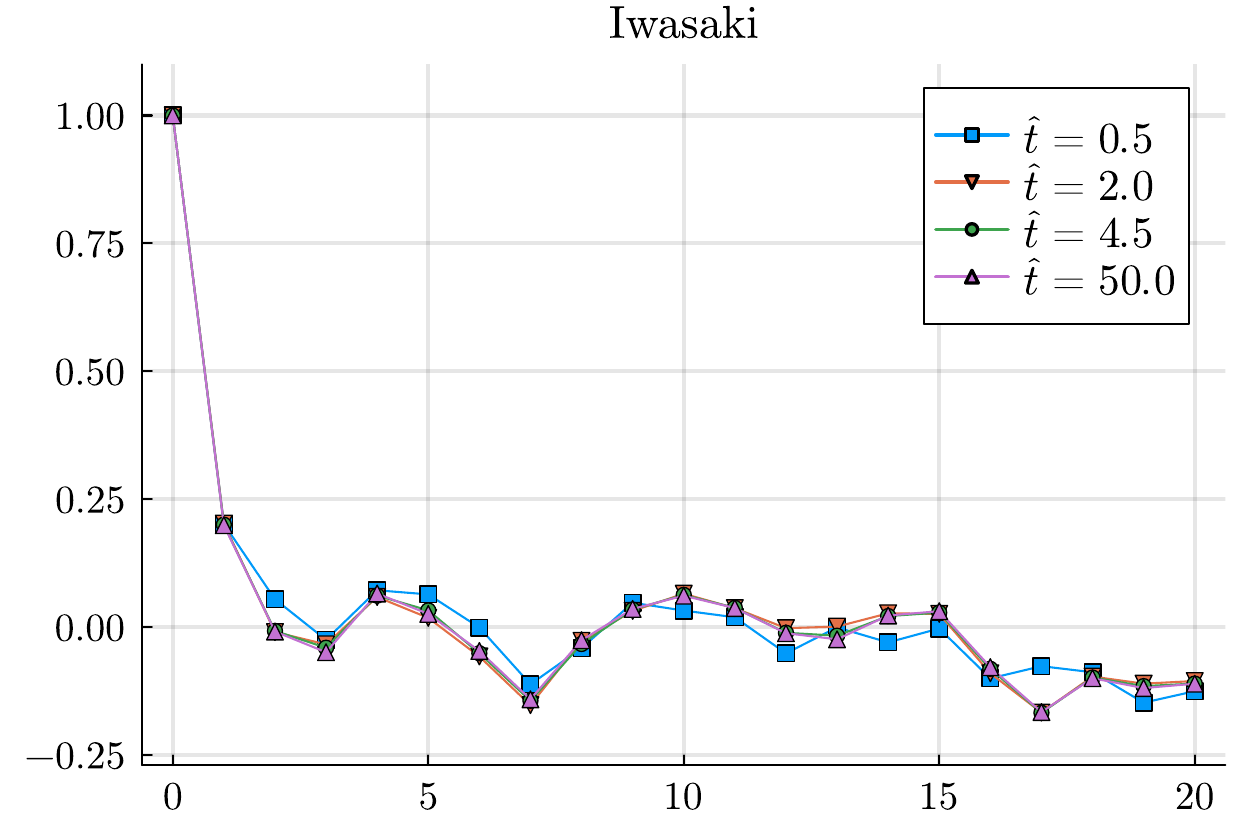}
\includegraphics[width=0.425\textwidth]{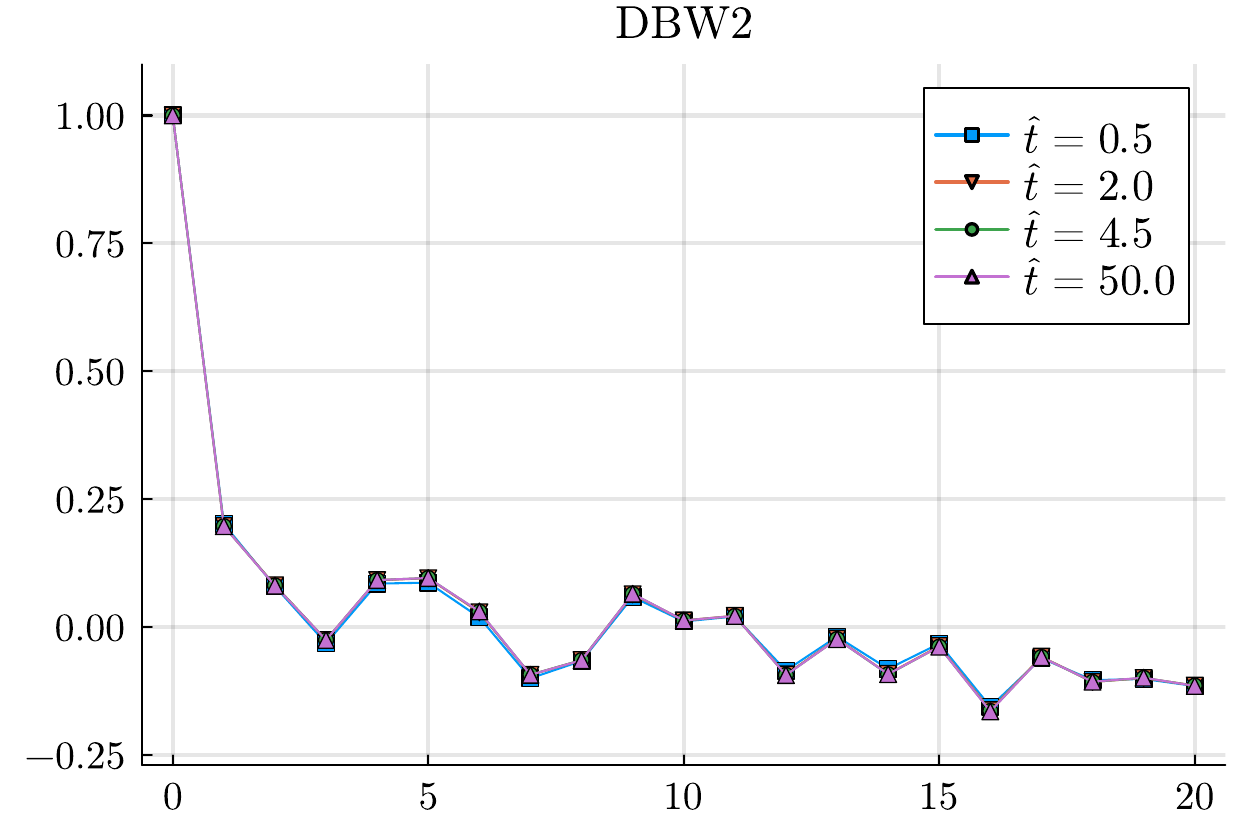}
\end{center}
\caption{
Autocorrelation functions for smeared $Q_\mathrm{top}$ by different gradient flows at flow time $\hat{t} = 0.5, 2.0, 4.5, 50.0$:~{\it (Upper-left)} Wilson, {\it (Upper-right)} Symanzik, {\it (Lower-left)} Iwasaki, {\it (Lower-right)} DBW2. The horizontal axis corresponds to the Monte-Carlo time $t'_\mathrm{MC} = t_\mathrm{MC}/20$ since gauge configurations are stored every 20 updates. 
\label{fig:acfunc_gf}}
\end{figure}

We measure the topological charge on the fly without flow.
The histogram is shown in Figure~\ref {fig:top-hist}. 
The topological charge fluctuates from less than $-4$ to $4$ in the production run.
One can see that the topological charge is correctly distributed in the production run from the histogram.

The autocorrelation function for the topological charge at flow time $\hat{t}=0$ is shown in Figure~\ref{fig:acfunc}.
The horizontal axis corresponds to the Monte-Carlo time.
The integrated autocorrelation time is $\tau_{\rm int}\simeq 1.4$.
We save every 20 configurations in the production run ahead of measurements.
We measure all of the observables using saved configurations.

To confirm if the autocorrelation is well suppressed for those saved configurations, Figure~\ref{fig:acfunc_gf} illustrates the autocorrelation function for the smeared topological charges by different gradient flows at flow time $\hat{t} = 0.5, 2.0, 4.5, 50.0$. The results of different flows behave roughly in the same manner.

\section{Diffusion length for over-improved actions}
\label{app:diffusion_length}

In this appendix, we discuss the diffusion length of the gradient flows with different choices of $c_1$. 
The goal is to confirm that those gradient flows equally work well for the purpose of smearing gauge fields at the perturbative level. In the following, we set $a=1$. 

The free part of the lattice action~\eqref{eq:FlowAction} with including the gauge-fixing term $S_{\mathrm{g.f.}}$ is given by~\cite{Weisz:1982zw,Iwasaki:1983iya} 
\begin{equation}
    S^{(0)}_{\mathrm{flow}} + S_{\mathrm{g.f.}}^{(0)}
    =
    \frac{1}{2}
    \int_{-\pi}^{\pi} \frac{\dd^4 k}{(2\pi)^4}
    \sum_{\mu\nu}\sum_{i=1}^{N^2-1}
    A_\mu^i(k)\qty[G_{\mu\nu}(k)-\qty(1-\frac{1}{\alpha})\khat_\mu\khat_\nu]A_\nu^i(-k), 
\end{equation}
where 
\begin{equation}
    G_{\mu\nu}(k)
    =
    \delta_{\mu\nu}\khat^2
    -
    c_1 \qty[
    (\khat^2 \khat_\mu^2 + \khat^4)\delta_{\mu\nu}
    -
    \khat_\mu\khat_\nu(\khat_\mu^2+\khat_\nu^2)
    ],
\end{equation}
and the lattice momentum is defined by
\begin{equation}
    \khat_\mu = 2\sin\frac{k_\mu}{2},
    \qquad
    \khat^2 = \sum_\mu {\khat_\mu}^2,
    \qquad
    \khat^4 = \sum_\mu\khat_\mu^4.
\end{equation}
Here, note that $\hat{k}^4\not=(\hat{k}^2)^2$ in this notation.
Taking the lattice Feynman gauge $\alpha=1$, the flow equation in the momentum space is
\begin{align}
    \partial_t B_\mu(t,k)
    & =
    - \sum_{\nu} G_{\mu\nu}(k) B_\nu(t,k) \notag\\
    &= 
    -\qty[
        \khat^2 - c_1 (\khat^2\khat_\mu^2+\khat^4)
    ]B_\mu(t,k) - c_1 \hat{k}_\mu \sum_{\nu} \hat{k}_\nu (\hat{k}_\mu^2+\hat{k}_\nu^2)B_{\nu}(t,k).
\end{align}

\begin{figure}[t]
    \centering
    \includegraphics[width=0.45\linewidth]{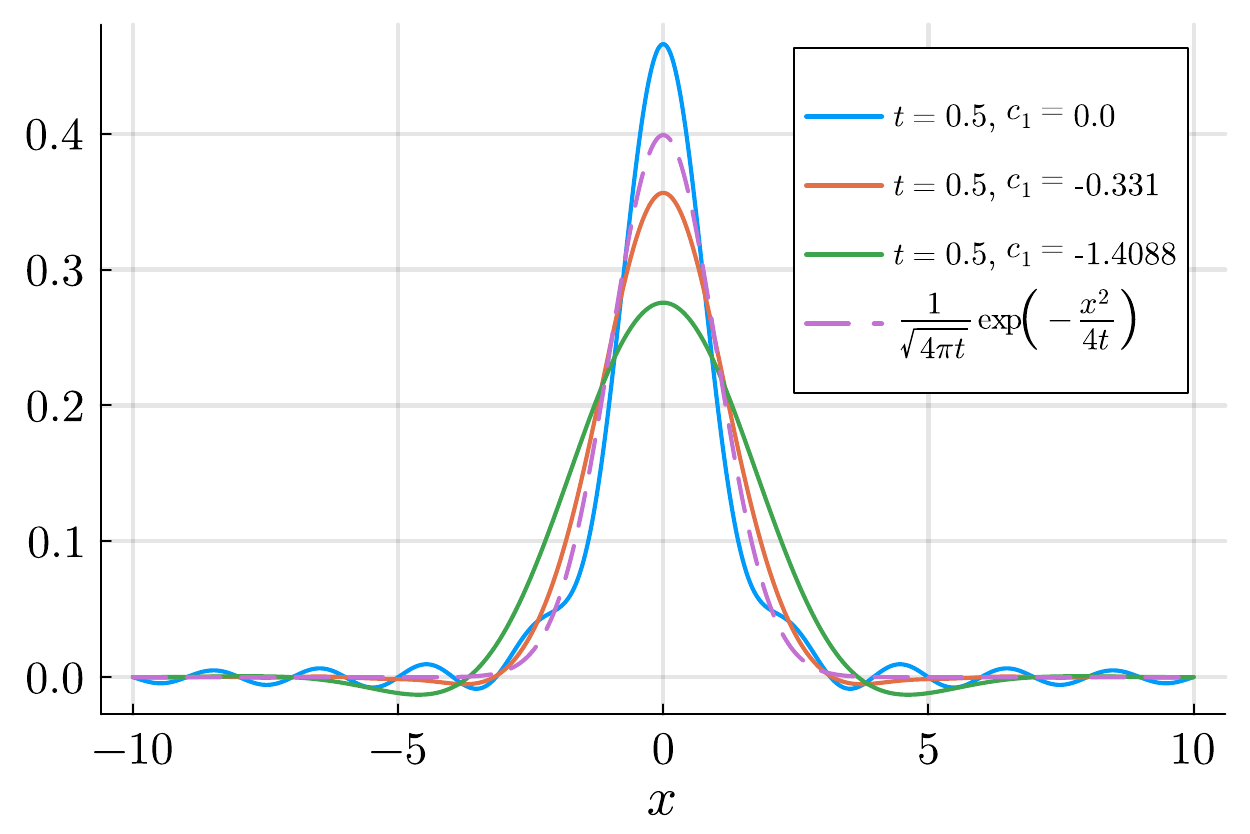}
    \includegraphics[width=0.45\linewidth]{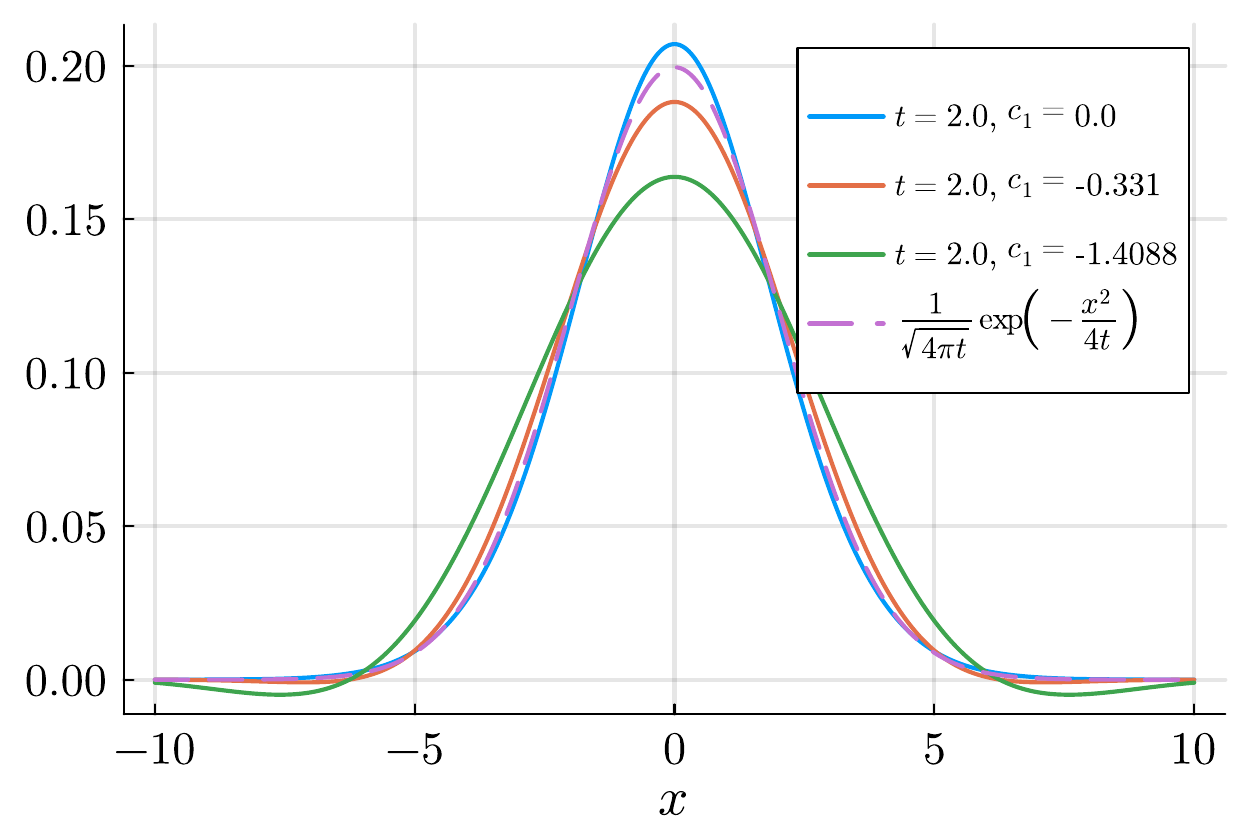}
    \caption{Plot of 1D heat kernel~\eqref{eq:GF1D} with $a=1$ for various $c_1$. [Left]~flow time $t=0.5$. [Right]~flow time $t=2.0$. The dashed line shows the analytic form of the Gaussian heat kernel as a reference.}
    \label{fig:GF1D}
\end{figure}

For simplicity, let us set $k_\mu=(k_0,0,0,k_3)$ so that we can easily diagonalze $G_{\mu\nu}(k)$, which means that the initial gauge field $A_\mu(x)$ is constant along the $1$-,$2$-directions. The eigenvalues of $G_{\mu\nu}(k)$ in this setting are given by 
\begin{align}
    \left\{\begin{array}{ll}
        \hat{k}^2-c_1 \hat{k}^4 \quad & \text{(2 transverse modes)},\\
        \hat{k}^2-c_1 (\hat{k}^2)^2 \quad & \text{(longitudinal mode)},\\
        \hat{k}^2 & \text{(gauge mode)}. 
    \end{array}
    \right.
\end{align}
We note that the eigenvalues take a more complicated form for generic $k_\mu$. Here, we are just trying to estimate the effect of $c_1$ on the diffusion length, so let us work on the transverse modes in this simplified setup neglecting the $1$-,$2$-directions. 
Then, the heat kernel for the diffusion equation of the transverse modes is reduced to the $1$D integration, 
\begin{equation}
    K_{\rm 1D}(t,x)
    =
    \int_{-\pi}^{\pi} \frac{\dd k}{2\pi}
    \exp(-t (\khat^2 - c_1 \khat^4) + \im k x).
    \label{eq:GF1D}
\end{equation}
We perform this integral numerically and illustrate the distribution with respect to 1D coordinate $x$.
Fig.~\ref{fig:GF1D} shows the plot of $K_{\rm 1D}(t;x)$ obtained by numerical integration, and the result is compared with the Gaussian heat kernel $\frac{1}{\sqrt{4\pi t}}\exp(-\frac{x^2}{4t})$ with changing the value of $c_1$ from $c_1 = 0$ (: Wilson) to $c_1 = -0.331$ (: Iwasaki) or $c_1 = -1.4088$ (: DBW2). 
While there is a tendency that the distribution becomes wider by making $c_1$ larger negative values, it seems to be mild enough so that the standard estimate of the diffusion length, $\sqrt{8t}$, is applicable even when we use the RG-improved (or over-improved) lattice actions for the flow equation.

\bibliographystyle{utphys}
\bibliography{./ref.bib, ./references_additional.bib}

\end{document}